\documentclass[%
 aip,pop,
 amsmath,amssymb,
 reprint,%
author-numerical,%
]{revtex4-1}

\usepackage{graphicx}
\usepackage{dcolumn}
\usepackage{bm}

\usepackage[utf8]{inputenc}
\usepackage[T1]{fontenc}
\usepackage{multirow}
\usepackage{mathptmx}
\usepackage{etoolbox}
\usepackage{comment}
\usepackage{xcolor}
\usepackage{soul}
\usepackage{enumitem}
\usepackage{siunitx} 
\usepackage{tabularx}

\DeclareSIUnit{\bar}{bar}

\makeatletter
\def\@email#1#2{%
 \endgroup
 \patchcmd{\titleblock@produce}
  {\frontmatter@RRAPformat}
  {\frontmatter@RRAPformat{\produce@RRAP{*#1\href{mailto:#2}{#2}}}\frontmatter@RRAPformat}
  {}{}
}%

\makeatother
\renewcommand{\selectlanguage}[1]{} 

\begin{document}

\preprint{AIP/123-QED}

\title[Technical Status Report on Plasma Components and Systems]{Technical Status Report on Plasma Components and Systems\\in the context of EuPRAXIA}
\author{A. Biagioni}
\affiliation{ 
Laboratori Nazionali di Frascati, Via Enrico Fermi 54, 00044 Frascati, Italy
}%
\author{N. Bourgeois}
\affiliation{ 
Central Laser Facility, STFC Rutherford Appleton Laboratory, Didcot OX11 0QX, United Kingdom
}%
\author{F. Brandi}
\affiliation{ 
Intense Laser Irradiation Laboratory (ILIL), Istituto Nazionale di Ottica – Consiglio Nazionale delle Ricerche (CNR-INO), Sede Secondaria di Pisa, Via Moruzzi, 1, 56124 Pisa, Italy
}
\author{K. Cassou}
\affiliation{ 
Laboratoire de Physique des 2 Infinis Ir\`ene Joliot-Curie - IJCLab - UMR9012, CNRS,Universit\'e Paris-Saclay  - B\^at. 100 - 15 rue
Georges Cl\'emenceau 91405 Orsay cedex - France.
}
\author{L. Corner}
\affiliation{ 
Cockcroft Institute of Accelerator Science, University of Liverpool, Liverpool L69 3GH, United Kingdom
}
\author{L. Crincoli}
\affiliation{ 
Laboratori Nazionali di Frascati, Via Enrico Fermi 54, 00044 Frascati, Italy
}%
\author{B. Cros}
\affiliation{ 
Laboratoire de Physique des Gaz et des Plasmas - LPGP - UMR 8578, CNRS, Universit\'e Paris-Saclay, 91405 Orsay, France
}
\author{S. Dobosz Dufr\'enoy}
\affiliation{Universit\'e Paris Saclay, CEA , LIDYL, 91191 Gif sur Yvette, France}
\author{D. Douillet}
\affiliation{ 
Laboratoire de Physique des 2 Infinis Ir\`ene Joliot-Curie - IJCLab - UMR9012, CNRS,Universit\'e Paris-Saclay  - B\^at. 100 - 15 rue
Georges Cl\'emenceau 91405 Orsay cedex - France.
}
\author{P. Drobniak}
\affiliation{ 
Department of Physics, University of Oslo, 0316 Oslo, Norway
}
\author{J. Faure}
\affiliation{ 
LOA, CNRS, École Polytechnique, ENSTA Paris, Institut Polytechnique de Paris, Palaiseau, France
}
\author{G. Gatti}
\affiliation{
Centro de Laseres Pulsados (CLPU), Edificio M5. Parque Científico. C/ Adaja, 8. 37185 Villamayor, Salamanca, Spain}
\author{G. Grittani}
\author{S. Lorenz}
\affiliation{ 
ELI Beamlines Facility, Extreme Light infrastructure ERIC, 252 41 Dolni Brezany, Czech Republic
}
\author{H. Jones}
\affiliation{ 
Cockcroft Institute of Accelerator Science, University of Liverpool, Liverpool L69 3GH, United Kingdom
}
\author{B. Lucas}
\affiliation{ 
Laboratoire de Physique des 2 Infinis Ir\`ene Joliot-Curie - IJCLab - UMR9012, CNRS,Universit\'e Paris-Saclay  - B\^at. 100 - 15 rue
Georges Cl\'emenceau 91405 Orsay cedex - France.
}
\author{F. Massimo}
\affiliation{ 
Laboratoire de Physique des Gaz et des Plasmas - LPGP - UMR 8578, CNRS, Universit\'e Paris-Saclay, 91405 Orsay, France
}
\author{B. Mercier}
\affiliation{ 
Laboratoire de Physique des 2 Infinis Ir\`ene Joliot-Curie - IJCLab - UMR9012, CNRS,Universit\'e Paris-Saclay  - B\^at. 100 - 15 rue
Georges Cl\'emenceau 91405 Orsay cedex - France.
}
\author{A. Molodozhentsev}
\affiliation{ 
ELI Beamlines Facility, Extreme Light infrastructure ERIC, 252 41 Dolni Brezany, Czech Republic
}
\author{J. Monzac}
\affiliation{ 
LOA, CNRS, École Polytechnique, ENSTA Paris, Institut Polytechnique de Paris, Palaiseau, France
}
\author{R. Pattathil}
\affiliation{ 
Central Laser Facility, STFC Rutherford Appleton Laboratory, Didcot OX11 0QX, United Kingdom
}%
\author{G. Sarri}
\affiliation{ 
School of Mathematics and Physics, Queen’s University Belfast, Belfast, UK
}
\author{P. Sasorov}
\affiliation{ 
ELI Beamlines Facility, Extreme Light infrastructure ERIC, 252 41 Dolni Brezany, Czech Republic
}
\author{R. J. Shalloo}
\affiliation{ 
Deutsches Elektronen-Synchrotron DESY, 22607 Hamburg, Germany
}
\author{L. Steyn}
\affiliation{ 
Laboratoire de Physique des Gaz et des Plasmas - LPGP - UMR 8578, CNRS, Universit\'e Paris-Saclay, 91405 Orsay, France
}
\author{M. Streeter}
\affiliation{ 
School of Mathematics and Physics, Queen’s University Belfast, Belfast, UK
}
\author{D. Symes}
\affiliation{ 
Central Laser Facility, STFC Rutherford Appleton Laboratory, Didcot OX11 0QX, United Kingdom
}%
\author{C. Thaury}
\affiliation{ 
LOA, CNRS, École Polytechnique, ENSTA Paris, Institut Polytechnique de Paris, Palaiseau, France
}
\author{A. Vernier}
\affiliation{ 
LOA, CNRS, École Polytechnique, ENSTA Paris, Institut Polytechnique de Paris, Palaiseau, France
}
\author{J. C. Wood}
\affiliation{ 
Deutsches Elektronen-Synchrotron DESY, 22607 Hamburg, Germany
}

\date{\today}


\maketitle

\begin{quotation}
The EuPRAXIA project \cite{Walker2017} aims to construct two state-of-the-art accelerator facilities based on plasma accelerator technology. Plasma-based accelerators offer the possibility of a significant reduction in facility size and cost savings over current radio frequency (RF) accelerators. The two facilities - one laser-driven one a beam-driven - are envisioned to provide electron beams with an energy in the range of 1-5 GeV and beam quality comparable to existing RF machines. This will enable a versatile portfolio of applications from compact free-electron laser (FEL) drivers to sources for medical and industrial imaging.

At the heart of both facilities is the use of plasma-based accelerator components and systems which encompass not only the accelerating medium itself, but also a range of auxiliary systems such as plasma-based electron beam optics and plasma-based mirrors for high-intensity lasers. From a technical standpoint, a high-degree of control over these plasma devices will be essential for EuPRAXIA to achieve its target performance goals. The ability to diagnose and characterize these plasma devices and to simulate their operation will be further essential success factors. Additionally, compatibility with extended operation at high-repetition rates and integration into the accelerator beamline will also prove crucial. 

In this work, we aim to review the current status of plasma components and related systems for both laser-driven and beam-driven plasma accelerators and to assess challenges to be addressed regarding implementation at future EuPRAXIA facilities.
\end{quotation}

\section{\label{sec:intro}Introduction}
With regards to plasma components and systems relevant for EuPRAXIA, the plasma within these devices is generated either via a high-voltage discharge or by a suitable laser pulse / particle beam. The generation of the plasma can be initiated from a few microseconds before up to just prior to the desired interaction. The neutral gas required for a plasma device is is typically fed into an interaction region via inlets or through an orifice; this interaction region can be effectively boundless (interaction occurs in gas flow as it freely enters a large vacuum chamber from an orifice) or confined (interaction occurs in a device which limits gas conductance into the surrounding vacuum environment of the accelerator). In a small number of cases the plasma is generated not from a gas but rather a solid / liquid surface. 

\subsection{\label{sec:plasma acceleration} Plasma Acceleration}
An accelerating structure can be generated in a plasma by creating a separation of charge. In plasma-based accelerators this charge separation is usually generated either by an intense laser pulse or by a high-current particle beam (known as the drive beam) which acts to displace the plasma electrons from its path while the plasma ions remain essentially stationary on short timescales. In the laser driven case the plasma electrons are displaced by the ponderomotive force which pushes electrons away from areas of high laser intensity. In the beam driver case, the plasma electrons are displaced by the Coulomb force. In both cases, for the regimes of interest for EuPRAXIA, the driver width and length must be matched to the scale of the plasma structure which varies with the plasma wavelength $\lambda_p = \sqrt{\pi / r_e n_e}$ , where $r_e$ is the classical electron radius and $n_e$ is the electron density. From this, one can also derive the natural frequency of plasma oscillations $\omega_p = 2 \pi c/ \lambda_p$. In addition to matching the scale of the plasma, the driver must additionally fulfill certain strength requirements; for laser drivers, the intensity of the pulse must be high enough such that free electrons in the laser field start to oscillate relativistically, for beam drivers the density of the drive beam must significantly exceed that of the background plasma density. These driver constraints lead to two different density operating regimes. In the case of commonly used Ti:Sa laser drivers, reaching relativistic intensities requires compressing the drive laser pulse to the few or few tens of femtoseconds and focusing the resulting beam to a few or a few tens of microns. This corresponds to typical operating densities in the range of \SIrange[print-unity-mantissa=false]{1e17}{1e19}{\per\centi\meter\cubed}. In the beam driver case the compressed electron beam is typically of order one to a few hundred femtoseconds and must maintain a beam density higher than the background plasma density. This typically corresponds to plasma densities in the range of \SIrange[print-unity-mantissa=false]{1e14}{1e16}{\per\centi\meter\cubed}. The operating density further impacts the required length of the accelerator as the accelerating field scales with the density. For example in the 1D non-relativistic wave breaking limit, the maximum accelerating field that can be sustained by the plasma is given as $E_{\mathrm{WB}} = 96 \sqrt{n_e \mathrm{[ cm^{-3}]}}$. In this simplified picture, laser-driven accelerators operate with accelerating fields in the range \SIrange[print-unity-mantissa=false]{30}{300}{\giga\volt\per\meter} and beam-driven accelerators in the range \SIrange[print-unity-mantissa=false]{1}{10}{\giga\volt\per\meter}.

\subsection{\label{sec:Overview} Plasma Components and Systems}
As can be seen above, the plasma conditions strongly impact the operation of the plasma acceleration process.
This can also be seen in auxiliary plasma components and systems such as active plasma lenses --- in which a current pulse is driven through a plasma to generate strong azimuthal magnetic fields capable of focusing relativistic electron bunches --- and plasma mirrors --- in which an overdense plasma is created on the surface of a liquid or solid to reflect incident laser light.
Consequently, manipulating and tailoring the properties of plasma components and systems is essential to controlling these key processes.
Below we discuss the generation and control of plasma components and systems for a wide variety of tasks, including beam-driven and laser-driven plasma acceleration as well as electron and laser beam control.
Additionally, we explore material robustness, longevity, plasma component integration and sustainability. Finally we discuss the measurement and simulation of plasma components and systems.

\subsection{\label{sec:eupraxiaparameter} EuPRAXIA accelerators parameters}

EuPRAXIA European facilities forsee two sites, one based on beam-driven plasma acceleration and one based on laser-driven plasma acceleration. Here we look specifically at the beam parameters envisaged for FEL applications.

\textbf{EuPRAXIA beam-driven FEL:} This facility will be based at INFN-LNF and driven by the EuPRAXIA \@ SPARCLAB RF accelerator \cite{Villa2023}. It comprises a photoinjector, which generates a high-brightness electron beam, and an X-band linac, which accelerates that beam to high energy. The accelerator can be operated a in single bunch or multi bunch mode up to $400\,$Hz, delivering electrons with energies up to $500\,$MeV. A beam-driven plasma accelerating stage attached to the linac will boost the beam energy to $1\,$GeV. An active plasma lens is envisaged for beam capture after the accelerating stage. Plasma systems and components parameters relevant to a beam-driven FEL facility at this site are summarized in the table Tab. \ref{tab:eupraxia-pwfa}. 

\begin{table}[!ht]
    \centering
    \caption{EuPRAXIA \@SPARC\_LAB parameters extracted from Ref\cite{Assmann2020} and last update\cite{Villa2023}}
    \label{tab:eupraxia-pwfa}
\begin{tabularx}{\columnwidth}{|X|X|c|X|}
    \hline
        \textbf{Device} & \textbf{Parameter} & \textbf{Value} & \textbf{Unit} \\ \hline
        \multirow{3}{4em}{Plasma Accelerator Stage} & Energy Gain & 0.5 & GeV \\ 
        ~ & Length & $>50$ & cm \\ 
        ~ & Density  & $10^{15}$ - $10^{17}$ & cm$^{-3}$ \\ 
        ~ & Repetition Rate  & 10-100 & Hz \\ \hline
        \multirow{3}{4em}{Active Plasma Lens} & Strength & 1-5  & kT \\ 
        ~ & Length & 2-4 & cm \\ 
        ~ & Density &  $1-10 \times 10^{17}$ & cm$^{-3}$ \\ 
        ~ & Repetition Rate & 10-100 & Hz \\ \hline
\end{tabularx}
\end{table}

\textbf{EuPRAXIA laser-driven FEL:} The location of this facility is to be decided from a shortlist in 2025. As such, various configurations are considered to achieve $1-5$~GeV beams\cite{Assmann2020}. The first configuration consists of a low-energy laser-plasma injector (LPI-LE) generating $150-500\,$ MeV high brightness electron beams, which would then be injected in a laser-driven plasma accelerating stage (LPAS) to reach a few GeV final energy. 

\begin{table}[!ht]
    \centering
    \caption{EuPRAXIA laser-driven general parameters from CDR\cite{Assmann2020} and references within.}
    \label{tab:eupraxia-lwfa}
    \begin{tabularx}{\columnwidth}{|X|X|c|X|}
    \hline
        \textbf{Scheme} & \textbf{Parameter} & \textbf{Value} & \textbf{Unit} \\ \hline
        \multirow{3}{4em}{LPI-LE}  & Energy Gain & 0.25-0.5 & GeV\\ 
        ~ & Density  & $10^{18}$ - $10^{19}$ & cm$^{-3}$ \\ 
        ~ & Length  & 1-10  & mm \\         
        ~ & Repetition Rate  & 10-100 & Hz \\  \hline
        \multirow{3}{4em}{Active Plasma Lens}  & Strength & 1-5  & kT \\ 
        ~ & Density &  $1-10 \times 10^{17}$ & cm$^{-3}$ \\ 
        ~ & Length & 2-4 & cm \\ 
        ~ & Repetition Rate & 10-20 & Hz \\ \hline
        \multirow{3}{4em}{LPI-HE / LPAS}  & Energy Gain & 1-5 & GeV \\ 
        ~ & Density  & $10^{17}$ - $10^{18}$ & cm$^{-3}$ \\
        ~ & Length  & 10-20  & cm \\   
        ~ & Repetition Rate  & 10-100 & Hz \\  \hline
    \end{tabularx}
\end{table}

The second layout considers PW-class laser driver to drive a high energy laser-plasma injector (LPI-HE) in a single stage to reach $1-5$~GeV beam. The third possible combination is the hybrid configuration where LPI generates a bright beam driving a beam-driven accelerator in a second plasma stage. Like for the beam-driven site, active plasma lenses are of significant interest\cite{Ferranpousa2019} for the capture section or coupling the beam in the LPAS stage. General parameters for plasma systems and components for a laser-driven FEL facility at this site are summarized in the table Tab. \ref{tab:eupraxia-lwfa}. Assman \textit{et al.} detailed the various configurations of plasma accelerator linacs for FEL in Ref\cite{Assmann2020} (see Tab. 8.1 and Fig 8.3).  


\section{\label{sec:gastarget}Neutral gas target}

Gas targets are systems that deliver a finite volume of neutral gas with which the driver or an auxiliary laser / discharge can interact to generate a plasma.  Gas targets are typically based on cells or jets to generate and shape a spatially delimited gas density profile. 

The main difference between these two approaches is the gas flow regime. In the case of gas jets, gas is forced to expand at supersonic speed into a vacuum chamber. In general, the density profile is determined by the internal geometry of the nozzle, the backing pressure, the orientation of the jet with respect to the driver beam axis and the height above the jet where the driver strikes the gas. For gas cells, the gas is kept relatively confined by an enclosure with typically mm- up to cm-scale transverse and longitudinal dimensions. Holes in the enclosure allow for gas injection inlets, specialized pumping outlets and apertures that allow the laser and electron beam to enter/exit. The pressure gradient drives the gas flow, leading to transonic flow with the Mach number potentially exceeding unity at the beam/driver in and out apertures. The injection gas inlet diameters are, in general, larger than the beam/driver in and out apertures. 


\subsection{\label{ssec:gastarget-bckg}Background}


In addition to its relative simplicity and widespread availability the main advantage of the gas jet is the open access provided to gas/plasma. This open geometry also avoids placing material which could be damaged by the driver close to the accelerator axis. Additionally, it facilitates convenient integration of plasma density diagnostics such as those that rely on transverse interferometric techniques (see section \ref{sec:diags-laser}). 

Two major types of cell designs have been used in LWFA experiments: (i) a \textit{tank}-like (or \textit{slab}, or \textit{reservoir}) \cite{Pollock2011,Vargas2014, Aniculaesei2018} geometry with almost static gas on axis (except at the cell apertures) composed of one or more gas volumes with typically cm-scale transverse dimensions, with some designs including a method of varying the gas length up to $10\,$cm\cite{Audet2018,Aniculaesei2018, Picksley2023}; (ii) \textit{channel}-like design with gas flow on axis confined in a hundreds-of-micron wide channel structure \cite{Osterhoff2008,Kirchen2021,Drobniak2023a}. In both cases, the gas cell walls are only intended to confine the gas in a controlled manner and not to guide a laser beam. Nevertheless, the jet or the cell targets can be used as a gas supply system for waveguide structures, such as the Hydrodynamic Optical-Field-Ionized (HOFI) channels (see sec. \ref{sec:HOFI}) and the channel-type cell target can be used for capillary discharge plasma waveguides (see sec. \ref{sec:discharge}). 

For all neutral gas sources, measuring and monitoring the electron density profile and plasma species distribution are crucial. Unlike gas jets, where access to the plasma is easy, in the case of gas cells, depending on the design, observation of the plasma or probing by a probe beam (see section \ref{sec:diags}) is carried out through optical windows with the inlet and outlet regions often obscured. However, the advantage of gas cells lies in the possibility of implementing pressure transducers or gauges\cite{Messner2020,Picksley2021phd,Drobniak2023phd} to obtain accurate measurements on the static pressure at different points within the cell. 

For gas targets used with laser-driven or hybrid electron sources\cite{Kurz2021}, there are several methods by which electrons can be injected into the plasma wakefield. Open gas jet targets allow colliding lasers to initiate injection with varying geometries, as in Refs~\cite{Faure2006,Kotaki2008,Golovin2018}.  Non-collinear/counterlinear geometries, such as those employed in Refs~\cite{Faure2006,Kotaki2008}, become more difficult when using cells and almost impossible when using capillaries due to the reduced access to the gas target. Manipulating the density longitudinal profile by inserting obstacles~\cite{Schmid2010} into the flow or by using another ``machining'' laser pulse~\cite{Chien2005} is also facilitated in gas jets. This has proven to be particularly relevant for density downramp injection~\cite{Bulanov1998,Suk2001a}.

Short-length gas cells have been widely developed to for electron sources based on ionization injection \cite{Pak2010}. This involves use of a mixed gas consisting predominantly of a low-Z species (eg. hydrogen) doped in small concentrations with a high-Z species (eg. nitrogen). The selective ionization of inner-shell electrons of the doped gas, close to the laser peak intensity, injects electrons at a specific phase of the wakefield, promoting a more controlled and efficient injection process \cite{Chen2012, Pak2010}. The density distribution of the gas cell, where this interaction occurs, is crucial for optimizing the effectiveness of ionization injection and beam loading~\cite{Lee2016}. By carefully shaping the gas density profile and doping concentration, one can control the location and amount of injected charge, thereby tuning the beam characteristics to suit specific experimental or application requirements. The gas cells used for ionization injection are typically split into two sub-compartments. The laser first enters a zone of mixed gas where electron injection into the plasma cavity occurs (ionization injection) and then a second pure gas zone where previously trapped electrons are accelerated in the laser-driven plasma cavity \cite{Pollock2011}. This technique allows the decoupling of each process (ionization injection versus acceleration), as long as the first doped sub-compartment is controlled by mitigating the potential mixing at the interface between the two zones \cite{Kirchen2021,Drobniak2023phd}.

The gas target features depending on the gas delivery system and flow are summarized in the table Tab. \ref{tab:gastarget-features}
\begin{table}[!ht]
    \centering
    \caption{Current status of neutral gas target characteristics as a function of the gas delivery system. (*) Note that the lack of published results on high repetition rate operation with a cell-type target and a $>30\,$TW laser driver is due to the lack of availability of such a laser system. Gas jet studies at kHz are also limited to a few TW laser drivers and reported in the present paper with a few Watts of average power.}
    \label{tab:gastarget-features}
    \begin{tabular}{|l|c|c|l|}
    \hline
        \textbf{target type} & \textbf{jet} & \textbf{cell} & \textbf{unit} \\ \hline \hline 
        gas flow & supersonic & transonic & - \\ \hline
        access & open & limited & ~ \\ \hline
        length & $0.100-100$ & $2-100$ & mm \\ \hline
        density & $10^{18} - 10^{20}$ & $10^{17}-10^{19}$ & cm$^{-3}$ \\ \hline
        density control & $1$ & $0.3$ & \% \\ \hline
        repetition rate$^{(*)}$ & $1-1000$ & $1-10$ & Hz \\ \hline
        driver average power & $\sim 2$ & $\sim 2$ & W \\ \hline
        density tailoring & yes, sharp & yes, smooth & - \\ \hline
        species distribution  & yes & yes & -\\ \hline
        multi- driver beam  & yes, any angle & yes, co-linear & - \\ \hline
        gas load & high & moderate & ~ \\ \hline
        wear part & nozzle & aperture / channel & ~ \\ \hline
    \end{tabular}
\end{table}

\subsection{\label{ssec:gastarget-soa} State of the Art}
In the following, we review three aspects of gas target development that are relevant for EuPRAXIA: (i) the tailoring of the density profile for better control over processes such as electron injection and, more generally, for controlling the electron phase space distribution, (ii) the generation of long gas structures and (iii) recent developments relating to high repetition rate operation.

\subsubsection{Tailored density profiles}
Initial work on gas jet design for laser-driven plasma accelerators focused on the fact that cylindrical supersonic jets were advantageous for underdense laser-plasma interactions in general. In Ref.~\cite{Semushin2001}, it was shown that the divergence angle of the jet $\alpha$ scales as $\sin\alpha=1/M$ where $M$ is the Mach number. This indicates that supersonic jets with high Mach numbers provide a relatively collimated flow with short-density gradients at the entrance and exit of the jet, as well as a relatively flat-density plateau within the jet. 
In Ref.~\cite{Schmid2012}, Schmid and Veisz presented an extensive study and optimization of cylindrical gas jets, extending the discussion to the case of micro-nozzles with typical tip openings at the $100\,\mu$m level. Such small jets permit to produce high plasma densities exceeding $10^{20}\,$cm$^{-3}$ over $100\,\mu$m scales, which is useful for plasma accelerators driven by few-cycle laser pulses~\cite{Guenot2017,Faure2019,Salehi2020}.

\begin{figure}
	\centering
	\includegraphics[width=0.45\textwidth]{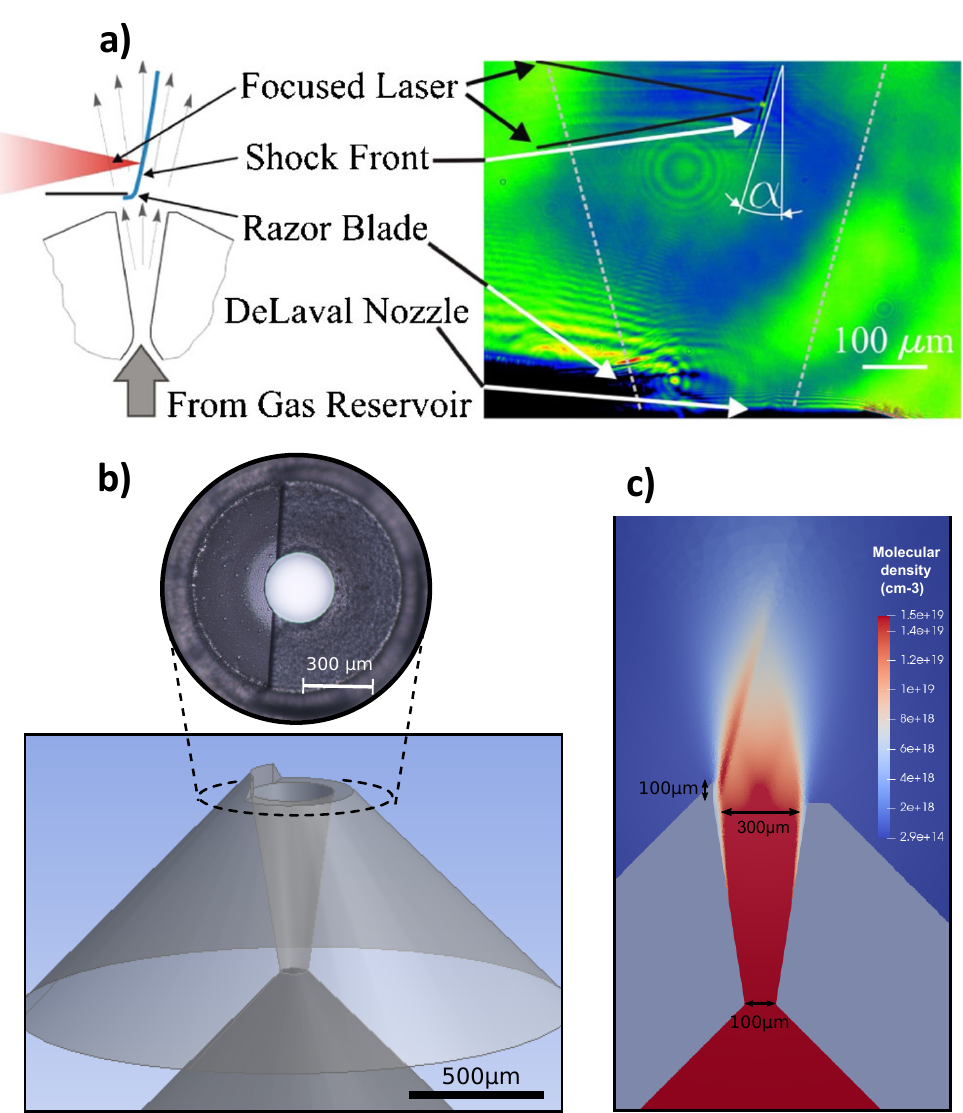}
	\caption{a) Left: Concept of shock front generation using a supersonic nozzle and a knife edge. Right: shadowgraphy image showing the shock. Images taken from Ref.~\cite{Schmid2010}. b) Concept of a one-sided shocked nozzle where the obstacle used to create the shock is directly embedded in the nozzle itself. c) CFD simulations for the nozzle geometry of b), clearly showing the emergence of a shock with 10~µm scale. Images taken from Ref.~\cite{Rovige2021}. }
    \label{fig:shock}
\end{figure}

Over the past two decades, progress has been made towards tailoring the density profile provided by cylindrical supersonic jets. In particular, density down ramp injection is a robust way to spatially localize injection~\cite{Bulanov1998,Suk2001a,Brantov2008} and potentially provide high quality and more stable electron beams. Initial works relied on using a secondary laser pulse to spatially modulate the plasma density before the arrival of the main pulse, thereby creating a density down-ramp for triggering electron injection~\cite{Chien2005,Faure2010}. More recent works on density down-ramp injection often use shock-fronts to form sharply structured longitudinal density profiles, see Schmid et al.~\cite{Schmid2010} as well as a non exhaustive list of references~\cite{Buck2013,Burza2013,Thaury2015,Swanson2017}. The shock is produced when the supersonic flow from the jet encounters an obstacle, usually, a knife edge, see fig.~\ref{fig:shock}a. Longitudinal modulation of the electron density using a knife-edge insertion was used to re-phase the electrons and overcome the dephasing limit~\cite{Guillaume2015}, but also to perform energy chirp compensation of the electron bunch longitudinal phase space~\cite{Dopp2018}. Recently, a knife-edge in a slightly different geometry was used to create a long exit density gradient in a gas jet to decrease the electron beam divergence~\cite{Chang2023}. Note that these \emph{shocked} density profiles cannot be achieved using gas cells since they require a supersonic flow and subsequent compression of that flow near the inserted obstacle. Inserting a knife-edge in the flow with good precision can prove difficult when the gas jet is scaled down to hundred-micron-scale dimensions. It was therefore proposed to integrate the shock formation into the design of the nozzle~\cite{Rovige2021} to offer a more compact, robust, and simple solution than the external knife-edge. Such a design is shown in fig.~\ref{fig:shock}b and c.

The gas density distribution achieved in a gas cell is fully determined by the geometry of the gas cell and the gas inlet conditions. The first constraint on the design concerns the cell apertures through which the driver and beam propagate : (i) The gas flow into the vacuum chamber has to be minimized to stop a large build-up of gas that can negatively affect the laser propagation before entering the cell. 
(ii) it is often desired to achieve specific density gradients on the laser axis, mainly an entrance density up-ramp that is as steep as possible and an exit density out-ramp that allows for adiabatic electron beam propagation and emittance control.

In addition to the cell entrance and exit gas density distribution, the gas density and composition inside the cell is a core topic. Whether for a \textit{tank}-like or \textit{channel} design, whether with one or multiple gases, one wants to ensure control over the longitudinal and transverse gas distribution. The dimension and position of gas inlets and gas outlets (aperture pumping but also specific additional pumping holes) play a major role in the design, and the overall selection of these parameters is impacted by the cell apertures.  For a single species gas cell, the density profile can be tailored by varying the shape of the aperture between two regions with a pressure differential\cite{Drobniak2023b}, or the inner channel diameter, and injection inlet sizes, orientation and shape \cite{Kim2021}. Gas cells\cite{Pollock2011,Li2012}, like combined gas jets\cite{Golovin2015}, can have a tailored longitudinal distribution of the gas species: by using multiple regions\cite{Pollock2011,Vargas2014, Drobniak2023b} or by inserting a pumping aperture to limit inter-diffusion of gas species \cite{Kirchen2021}. The selection of the cell geometry is complex, and Computational Fluid Dynamics (CFD) simulations \cite{Audet2018,Dickson2022,Drobniak2023a} help in the design efforts and in tuning the geometric constraints (see section \ref{ssec:tools-cdf}).
Basic Reynolds number calculation and CFD simulations allow for the assertion that the gas flow remains laminar, except for the interfaces with the vacuum where the gas expands to transonic up to supersonic velocities at relatively high cross sections (cm-scales) and thus can potentially be turbulent. In the injection pipes and within the cell, high velocities are mostly reached within very small cross sections, in the hundreds of microns range, thus allowing the flow to remain laminar. 
Less turbulence, in theory, should lead to improved electron beam quality and shot-to-shot stability \cite{kuschel2018, Vargas2014, Osterhoff2008}, since turbulent flows are more susceptible to produce shot-to-shot density fluctuations and thus electron beam parameter variations.

\begin{figure}[ht]
\centering
\includegraphics[width=0.5\textwidth]{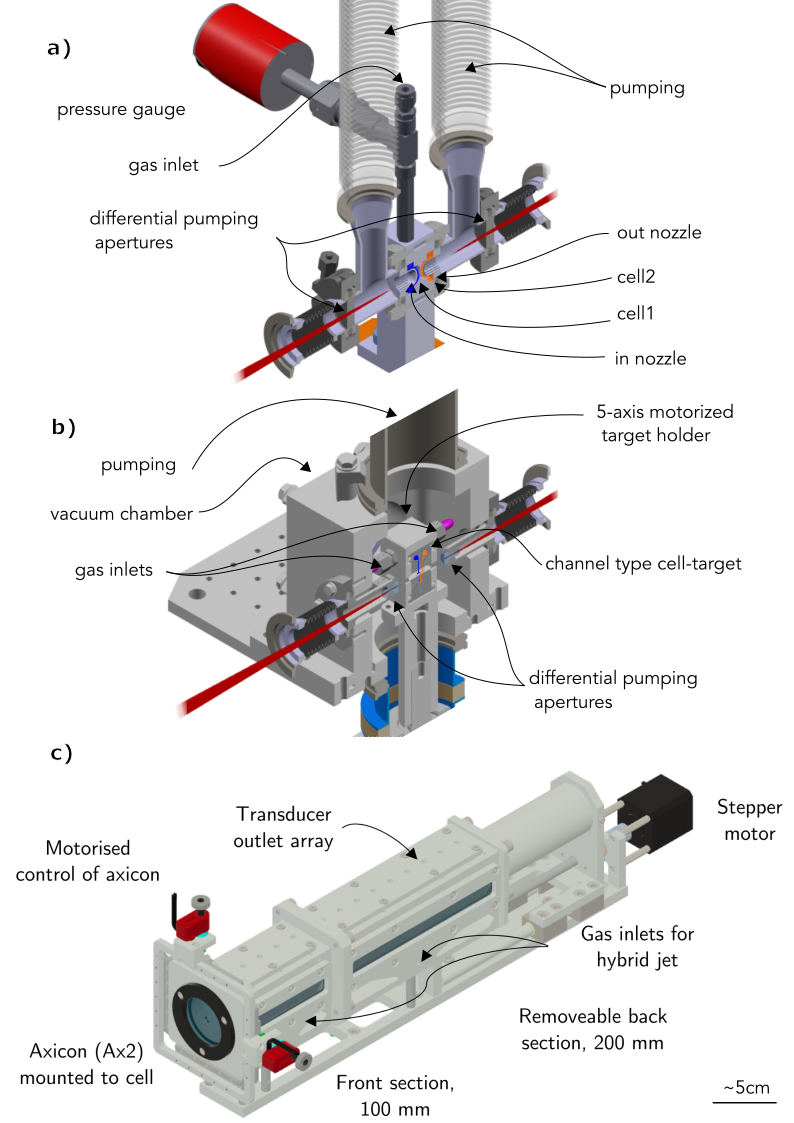}
\caption{ (a) - 3D section view of an example of a two-cell beamline integrated gas target prototype for ionization injection with ceramic input and output nozzles. The laser driver beam goes from left to right, and the left region length is $0.7\,$mm. The gas cell is operated in continuous-flow and coupled to two differential pumping crosses upward and downward \cite{Drobniak2023b}. (b) example of a channel-type gas cell for ionization injection for continuous flow operation similar to LUX design implementation\cite{Jalas2021}. A compact target chamber with differential pumping aperture can stand directly in the beamline (c) long gas cell for GeV range laser-driven plasma acceleration\cite{Aniculaesei2018,Picksley2023} for HOFI channel waveguide with front section closed by the axicon lens. 
}
\label{fig:gascell-twochamber}
\end{figure}

\begin{figure}[ht]
\centering
\includegraphics[width=0.5\textwidth]{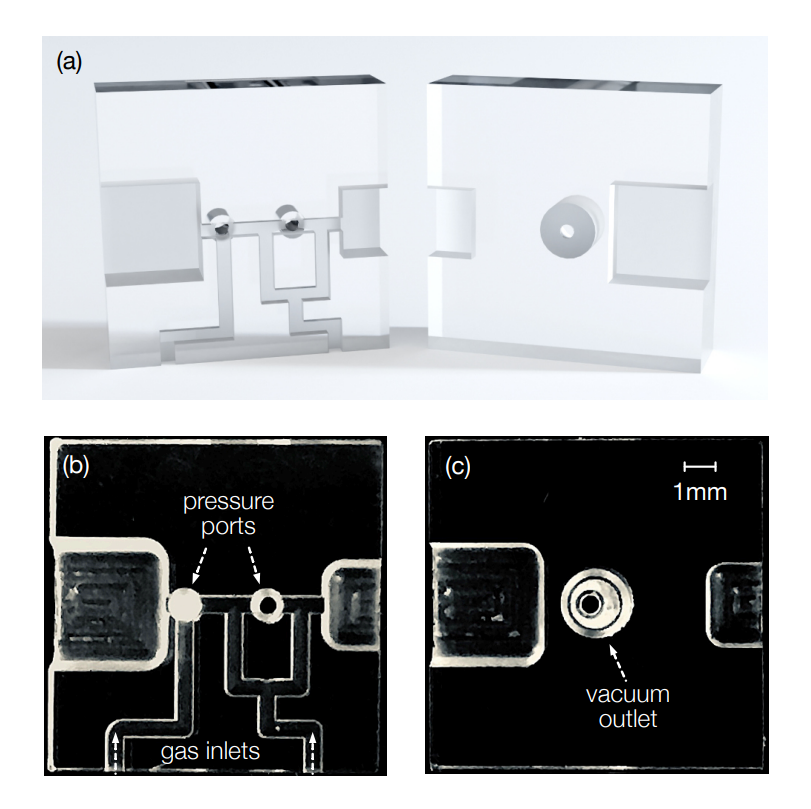}
\caption{An illustrative example of a channel-type gas cell for ionization injection made of two sapphire blocks with one gas inlet for a gas mixture and two more for pure hydrogen. Pressure ports are drilled for direct pressure measurement. A specific additional vacuum outlet decreases the diffusion of gas mixture \cite{Kirchen2021}.}
\label{fig:gascell-channel}
\end{figure}

The reservoir/tank/slab cells are generally assembled from many constituent parts --- eg a main body offering the sub-compartments, input and output nozzles for controlled gas flow etc. --- which allows the exchange of the parts damaged by the laser and ad-hoc customizations of the geometry of the cell (Fig.\ref{fig:gascell-twochamber}). 
In addition to easy maintenance, the motivation is also to offer a versatile design with geometrical tuning possibilities during operation. The channel cells are generally manufactured by etching a flow channel into a piece of glass, crystal, ceramic, or plastic material, with a second piece glued onto the etched piece to complete the assembly (Fig. \ref{fig:gascell-channel}). Alternatively, channel etching designs also exist with 3 layers instead of 2 \cite{Drobniak2023phd}, without milling of the surfaces, thus preserving optical quality for interferometric transverse optical diagnostics. As opposed to the reservoir/tank/slab design, the channel design does not offer, so far, any geometrical tunability during operation.


\begin{figure}[ht]
\centering
\includegraphics[width=0.5\textwidth]{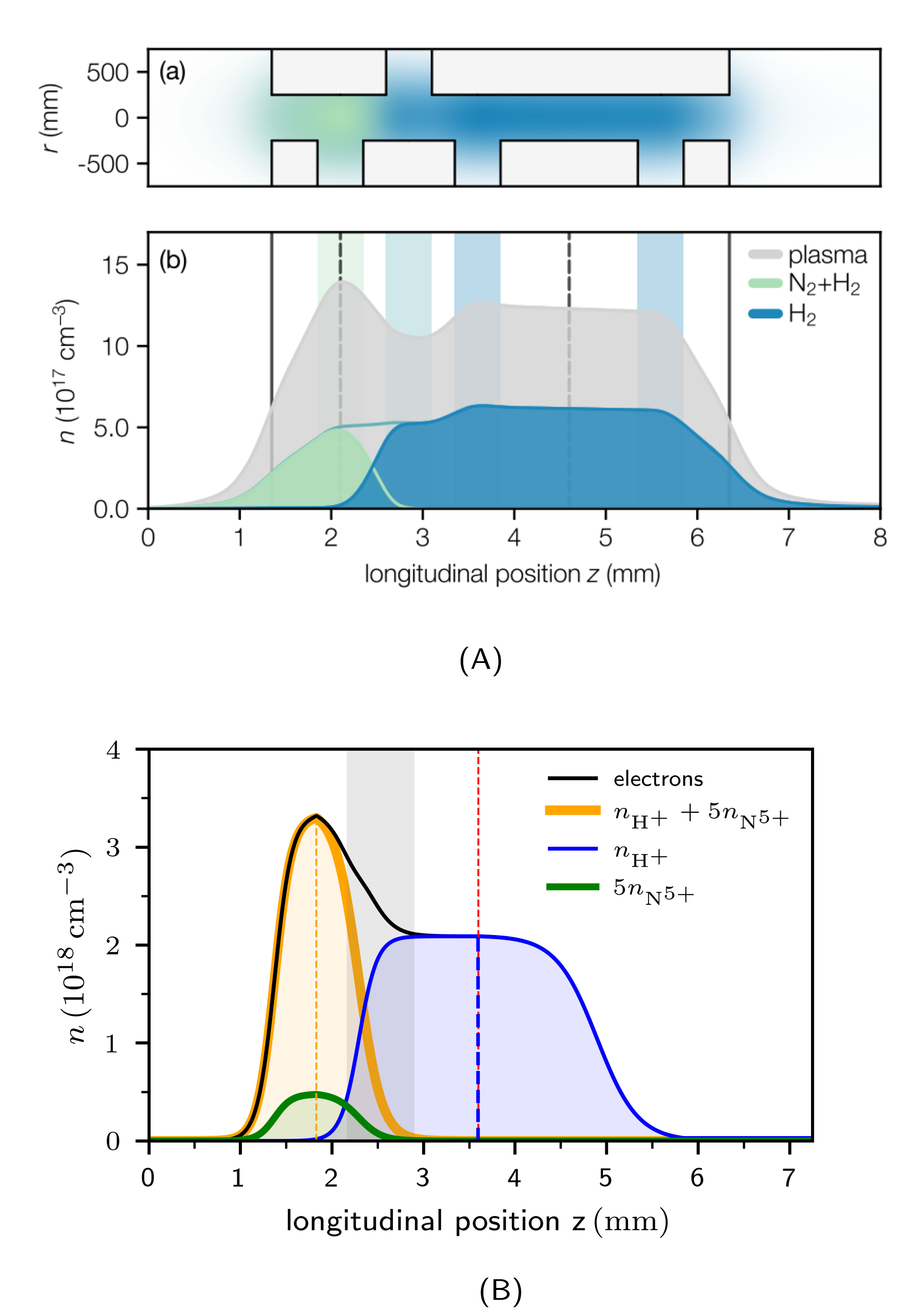}
\caption{(A) Figure from Ref.\cite{Kirchen2021phd} channel geometry with longitudinal gas and plasma density profile extracted from a CFD simulation. (A-a) Simplified representation of the channel layout. (A-b) Mixed gas (green), hydrogen (blue) and plasma electron (grey) density $n$. Location of the inlets and vacuum outlet (coloured bars), pressure ports (dashed lines) and interaction channel entrance and exit (solid lines). (B) An illustrative density profile obtained from numerical studies optimisation of a compact laser-plasma electron injector source shows nitrogen confinement in the first section and pure hydrogen in the second section of the target\cite{Marini2024}.}
\label{fig:gascells-densityprofile}
\end{figure}

The state-of-the-art gas cell configurations used for ionization injection-based electron sources comprise a two-region gas cell in a channel geometry with a first compartment of $0.5\,$mm length with 10\% nitrogen doping and a second compartment of $4\,$mm of pure hydrogen (see Fig. \ref{fig:gascells-densityprofile} for the density profile along the propagation direction). These sources have successfully demonstrated high shot-to-shot stability over $5000$ shots \cite{Kirchen2021}, the generation of beams with charges between $28$ and $60\,$pC and with average energies between $250$ and $300\,$MeV at energy spreads of $7\,$MeV (FWHM). 

\subsubsection{Long gas structure}
Extending the length of the gas/plasma structure from the few cm to meter scale is crucial for the LPI-HE scheme for EuPRAXIA. Slit nozzle gas jets from cm \cite{Grafenstein2023} to 30-cm in length have been developed \cite{Shrock2022, Picksley2024}. 
Variable length gas cells and gas cell tubes are used for PW-class experiments\cite{Aniculaesei2018, Picksley2023, Lee2024} or to provide a gas container for HOFI guiding structures \cite{Picksley2023} (see section \ref{sec:HOFI}).  

\subsubsection{High-repetition rate}

\begin{figure}
	\centering
	\includegraphics[width=0.45\textwidth]{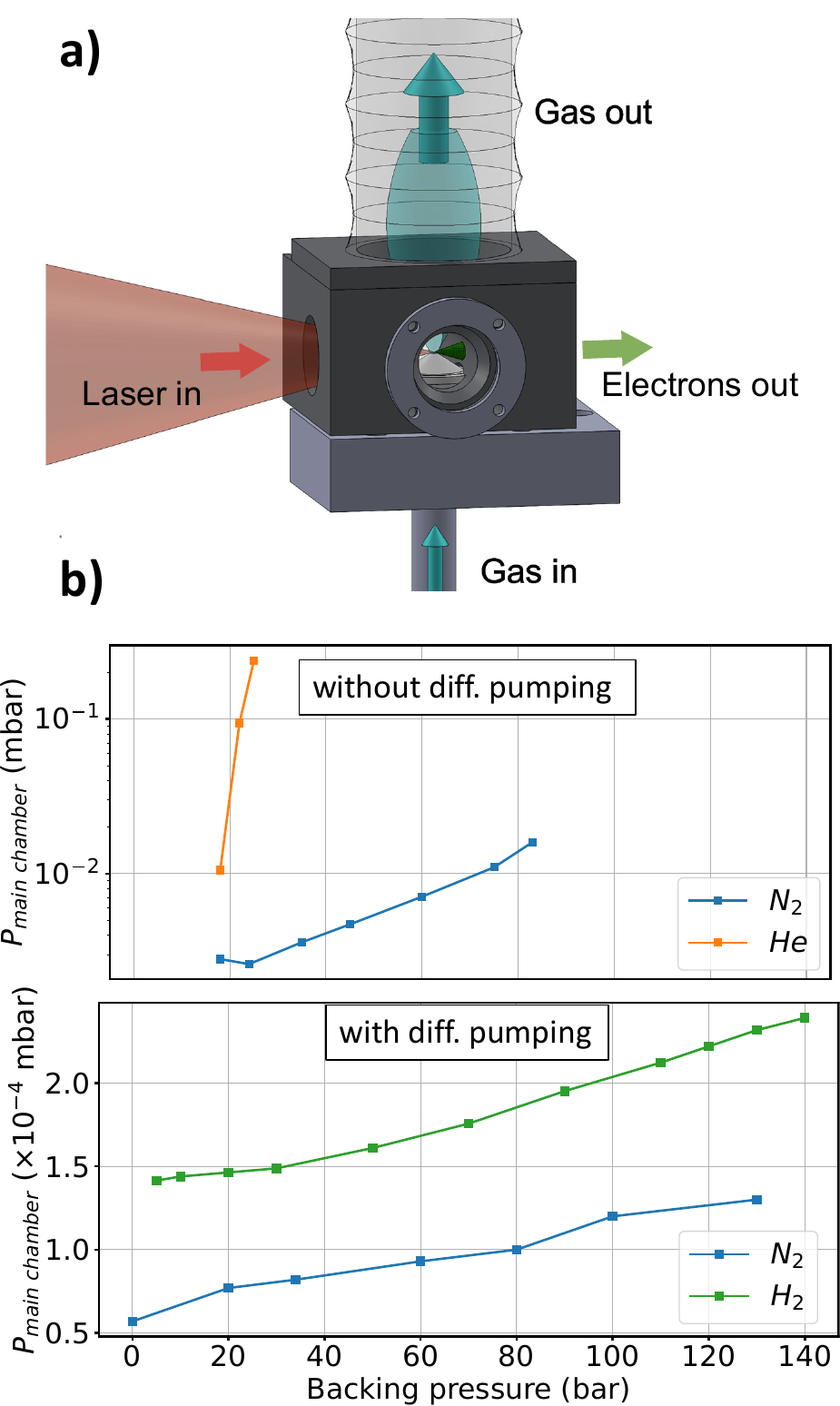}
	\caption{a) Conceptual drawing of differential pumping set-up for continuous flow operation: the gas jet is embedded in a small vacuum chamber which is pumped directly by a high-capacity primary pump that keeps the pressure in the small chamber at the bar level.  b) Measurements showing the residual pressure in the main chamber as a function of the backing pressure in the case where there is no differential pumping (top), and in the case with differential pumping (bottom). Images taken from Ref.~\cite{Monzac2024b}. }
    \label{fig:diffpump}
\end{figure}

Open gas jets are less prone to laser damage and, consequently, could be considered a strong candidate for future high-repetition rate laser-driven plasma accelerators. Application at high repetition rates requires the target to be able to produce the gas density profile repeatedly at the repetition rate of the driver. For this, there are two options for closed or open gas structures: pulsed flow or continuous flow. In a pulsed mode, the valve is only open for a limited time. Such valves typically have a time period of a few ms to 100s of ms, limiting in principle the repetition rate to the sub-kHz regime. However, until now, there has been no experimental evidence that a pulsed valve can produce the required plasma density at repetition rates beyond $10$~Hz for a laser-driven plasma accelerator. While kHz gas jets are routinely used for high harmonic generation~\cite{Even2014}, this process requires significantly lower density than laser plasma acceleration. On the other hand, in a continuous flow, the repetition rate for the jet or cell gas target is limited only by the driver or by the recovery time of the plasma itself \cite{DArcy2022}. However, the pumping system has to be carefully designed to avoid a detrimental stagnation pressure in the vacuum chamber. In Refs.~\cite{Guenot2017,Faure2019,Rovige2020}, the authors demonstrated the handling of a continuous flow gas jet for a kHz plasma accelerator using a nitrogen micro-jet and an electron plasma density in the range of $10^{20}\,$cm$^{-3}$. The pressure in the vacuum chamber was maintained at $10^{-3}$ mbar when operating in continuous flow, while most linear accelerators operate with vacuum levels at least as low as $10^{-5}$ mbar, requiring differential pumping (see sec. \ref{sec:integration-vac}). Indeed, the pressure is kept orders of magnitude lower to prevent collisions between the electrons and gas particles which will affect the quality of the electron beam when transported over longer distances than the typical meter-scale plasma accelerator experiment. Recent work has shown that implementing a differential pumping system for gas jets similar to gas targets for nuclear physics \cite{Shapira1985,Rapagnani2023,Brandi2020c}, see fig.\ref{fig:diffpump}a, permits maintaining a residual pressure in the chamber much lower than previously \cite{Monzac2024a}. It was shown that when using lighter gases such as hydrogen and helium at much higher backing pressures, $>100$ bar, the residual pressure could be maintained at the $10^{-4}$ level, see figure~\ref{fig:diffpump}b. This result highlights the fact that continuous operation of gas jets for plasma accelerators is now possible, even for light gases at high pressure. So far, this has only been demonstrated experimentally for $100\,\mu$m scale nozzles operating at high backing pressure, thus interpolating these results to longer jets with lower pressure should be considered. 

\noindent With gas cells providing less gas loading, it has been possible to operate them in continuous flow with differential pumping stages\cite{Maier2020, Kirchen2021, Drobniak2023b}. These works have shown that they are suitable for high repetition rate operation. 
Operation of gas cell structures in steady state flow allows precise control of the pressure, limited only by the sub-per cent accuracy of the mass-flow controller\cite{Messner2020}.
When operating in continuous flow with differential pumping, gas jet and gas cell systems tend to be similar, losing the advantage of the open access of the gas jet. However, in the case of jets, the gas supply is always provided by a supersonic jet with a preferential flow towards the gas catcher, which leaves the possibility of making multiple openings without significantly altering the gas density profile in the interaction region.  

The ablation of the entrance and exit apertures by the laser, and energy deposition in the channel-type cells are critical for high repetition rate operation. For the thermal loading inside the gas-cell target, recent work has investigated a design with liquid-cooling for $100\,$W operation for high harmonic generation\cite{Filus2022}. Nevertheless, the involved laser intensities are lower for high harmonic generation, limiting laser ablation issues.  



\subsection{\label{ssec:gastarget-chall}Challenges and future development}

Neutral gas target systems are the heart of any plasma components. The development of high repetition rate plasma structures based on supersonic jets or transonic gas circulation, depending on the density scale, density tailoring and species distribution required, is a major challenge for EuPRAXIA. With high-average power laser drivers ($\sim 100\,$W), reliable gas target designs have emerged. Extending this performance to a Joule-level laser system at a high repetition rates still remains to be tested and demonstrated. Research will have to address the issue of material resistance to laser and plasma damage (see sec. \ref{sec:materials}) for the next generation of high average power lasers, such as the ones available at the Extreme Light Infrastructure (ELI)~\cite{ELI} or proposed in the frame of LAPLACE~\cite{Laplace-HC}, KALDERA~\cite{Kaldera} and EuPRAXIA. Although major progress has been made in both numerical and experimental studies of gas cells for laser plasma acceleration\cite{Antipov2021,Drobniak2023a, Marini2024} significant challenges still remain. In the state-of-the-art case presented previously \cite{Kirchen2021}, 80\% of the shot-to-shot variations in beam loading were ascribed to laser fluctuations. The critical nature of understanding the laser variations and reconstruction of laser characteristics on every shot have been demonstrated \cite{Moulanier2023a,Moulanier2023b}.

The stability and quality of gas density \cite{kuschel2018, bohlen2022, lei2023} in the interaction region with the driver are critical for the high-quality beam and reliability required for EuPRAXIA plasma components.

Finally, closing the gas system loop (injection and pumping) by implementing gas recirculation would be one possible solution for reducing gas leakage and relaxing requirements on the differential pumping system. Additionally, this could improve the sustainable use of potentially scarce gases and reduce expenses.

\section{\label{sec:discharge}Capillary Discharge Sources}
Capillary discharge plasma sources typically rely on a channel-like gas cell geometry and typically comprise a sapphire or glass block through which a cylindrical channel, or capillary, is made. Neutral gas is supplied to this cylindrical channel via several gas inlets along the capillary. Plasma formation is initiated in these sources by means of a high-voltage electrical discharge between electrodes affixed to the ends of the capillary.


Gas-filled discharge capillaries can provide several diverse functions which can be of use to the plasma accelerator facilities envisaged for EuPRAXIA. The use of these sources as beam-driven plasma accelerator modules is well established ~\cite{pompili2024acceleration,Lindstrom2021,Lindstrom2024}. Additionally, they have found use as laser-driven plasma accelerator modules where the transverse structure of the plasma can be tuned to guide the drive laser pulse over distances commensurate with multi-GeV electron acceleration \cite{Leemans2006,Leemans2014,Gonsalves2019}. In such a context the devices are referred to as Capillary Discharge Waveguides (CDWs). Finally, the high amplitude azimuthal magnetic field generated by the current flowing along the capillary axis has found application in the focusing and deflection of charged particle beams ~\cite{Panofsky1950,pompili2024guiding,Lindstrom2018}. In this context they are referred to as Active Plasma Lenses (APLs).

\subsection{\label{sec:discharge-pwfa}Beam-Driven Accelerator Modules}
\subsubsection{\label{ssec:discharge-pwfa-bckg}Background}
\begin{figure}[ht]
\centering
\includegraphics[width=1.0\linewidth]{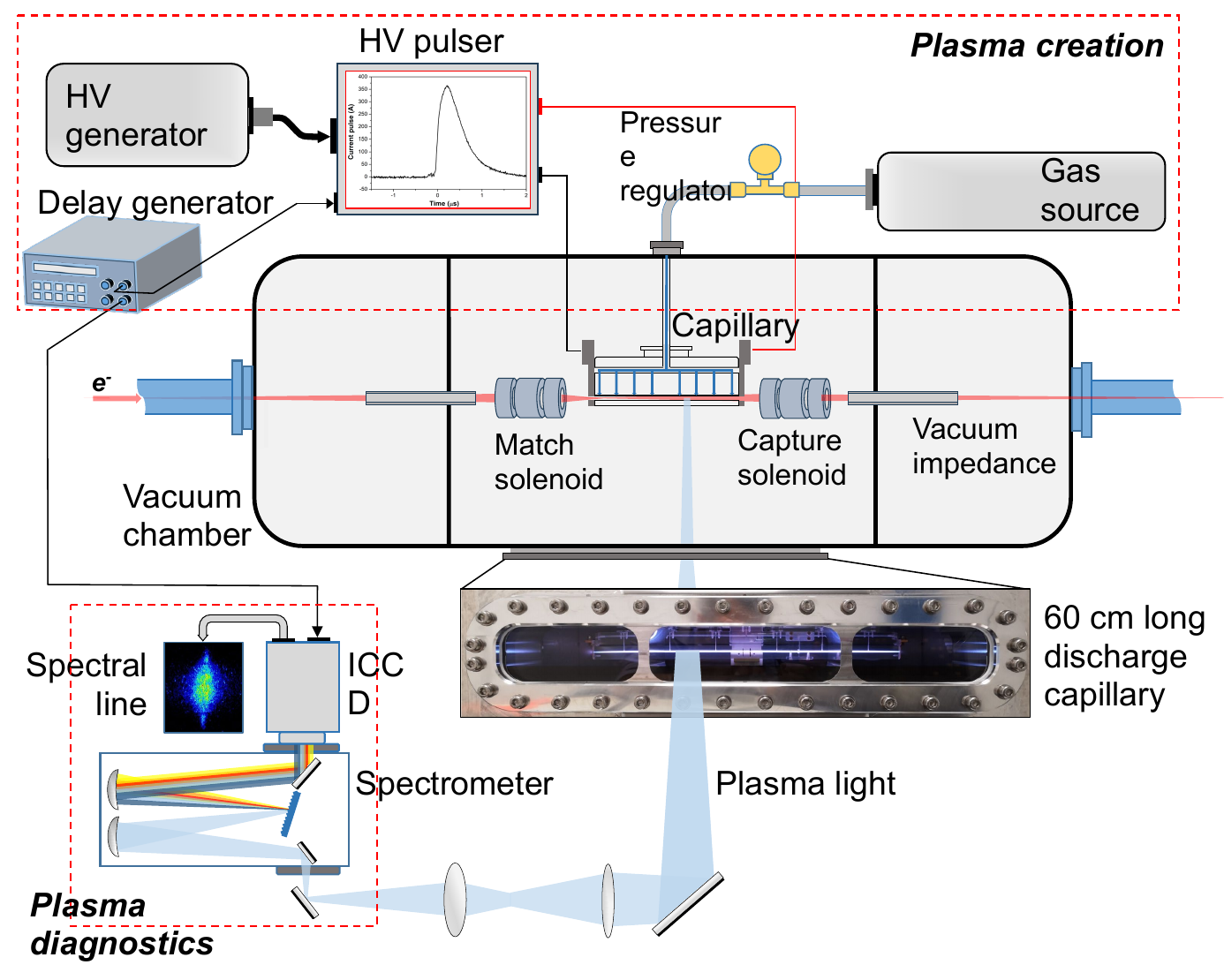}
\caption{General scheme of a plasma acceleration module based on gas-filled discharge capillaries. An image of 60 cm long gas-filled capillary-discharge is shown in this scheme. The plasma diagnostics is based on pressure broadening, for which the emitted plasma light is used to characterize plasmas.}
\label{scheme}
\end{figure}

Gas-filled capillary discharge sources are commonly used beam-driven accelerator modules at both the DESY FLASHForward \cite{DArcy2019} and INFN Frascati SPARC\_LAB facilities \cite{Ferrario2013}. 
Figure ~\ref{scheme} shows an example setup used to produce, control and characterize plasma for acceleration experiments. 
In recent years these capillary discharge plasma sources been integral to a number of developments in beam-driven plasma acceleration. Such developments include acceleration with high transfer efficiency \cite{Pena2024} and acceleration while retaining a small energy spread in the accelerated beam \cite{Lindstrom2021,Pompili2021}. More recently simultaneous emittance, charge and energy spread preservation has been demonstrated while operating with high-gradients and high efficiency \cite{Lindstrom2024}. A key result in illustrating the application of such beam driven plasm accelerators, and by extension the discharge plasma source technology upon which they rely, is the demonstration of free-electron lasing in a beam-driven plasma accelerator \cite{Pompili2022}. There have also been a series of studies investigating limits to the repetition rates of such sources \cite{DArcy2022,Pompili2024RepRate}.

\subsubsection{\label{ssec:discharge-pwfa-soa}State of the Art}

The capillary discharges used for plasma source formation are typically constructed of a hard material such as sapphire or diamond. 
They are on the scale of 10s to 100s of millimeters in length with diameters of hundreds of microns to a few millimeters.
The flow of gas into the capillary structure is often controlled via an electromechanical regulator or mass flow controller. 
A variety of pure and mixed neutral gases can be supplied to the capillary via the gas inlets. 
Some commonly used gases include hydrogen, nitrogen and argon. 
Gas pressures are typically in the range \SIrange{10}{100}{\milli\bar}, close to the minimum of the Paschen curve. The eventual density reached inside the source during acceleration can be lower than would be suggested by these backing pressures owing to the the fluid dynamics of the gas within the capillary and the evolution of the plasma during the discharge.
The electrical power required to create the high-voltage pulses for ionization of the gas column is provided by a high-voltage pulser circuit, which is powered by a high-voltage generator. 
The voltage pulses needed to produce the required plasma densities, which are in the range of $10^{15}$~cm$^{-3}$ up to $10^{17}$~cm$^{-3}$, are between a few and a few 10s of \si{\kilo\volt} with currents of the order of a few hundred amperes.
Typically the electrical pulse length is of order a few hundred nanoseconds to a microsecond and once formed the plasma decays on microsecond timescales. 
The complete scheme of a plasma accelerating module necessarily also includes a diagnostic system for the characterization of the plasma produced in the sources, which, in the case of capillary discharge, is often based on a spectroscopic technique using the pressure broadening effect for the measurement of the electron density distribution (see section \ref{sec:diags}).

In terms of high-repetition rate operation, kilohertz operation of water-cooled capillary discharges with plasma accelerator relevant parameters has been demonstrated~\cite{Gonsalves2016}. In addition to the drive electronics, the repetition rate of these capillary discharge systems is determined mainly by the following two physical effects~\cite{Sasorov2024}: i) averaged thermal balance of the capillary; and ii) recovering the pre-discharge gas distribution inside the capillary after the discharge. 

The thermal balance of the capillary is determined by the averaged energy deposition inside the capillary and the transport of this energy toward the outer boundary of the capillary and beyond. The energy deposited into the capillary at each cycle is determined by two sources; Ohmic heating of the gas discharge~\cite{Gonsalves2016} and energy deposition from the wake driver.

In addition to these practical considerations for extended operation at high-repetition rates, there have also been investigations into the maximum achievable repetition rate for beam-driven plasma acceleration stages \cite{DArcy2022,Pompili2024RepRate}.

\begin{figure}[ht]
\centering
\includegraphics[width=1.0\linewidth]{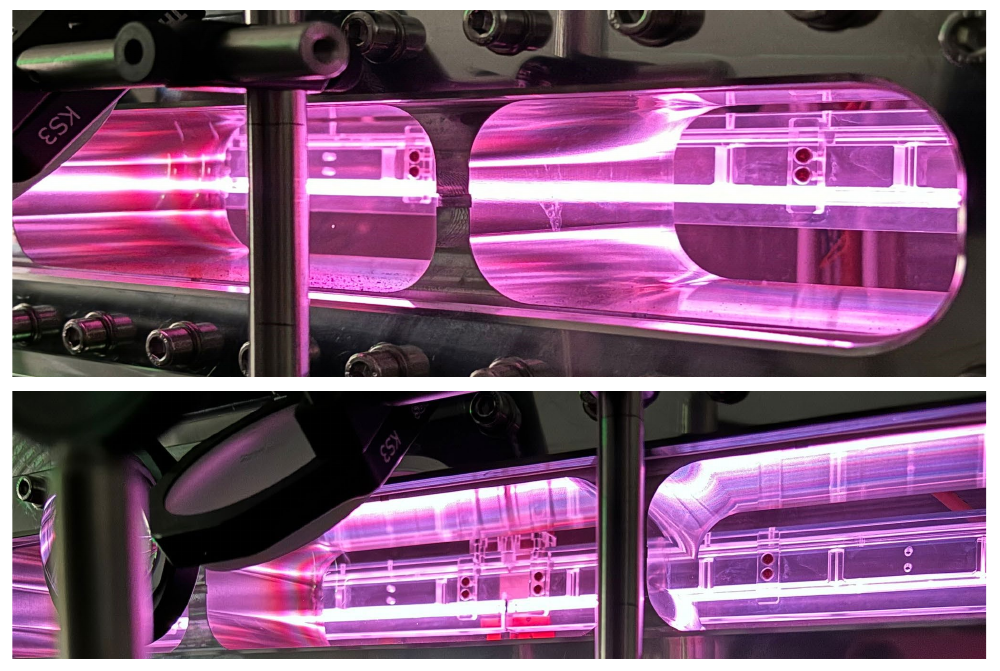}
\caption{High-voltage test discharges for plasma formation inside a 60-cm long source.}
\label{image60cm}
\end{figure}

The generation of long capillary-based plasma sources is an active area of research with questions such as gas supply, reliable discharging and optimization being investigated. 
At the SPARC\_LAB facility at INFN, the generation of \SI{60}{\centi\meter} capillary discharges have been demonstrated (fig.~\ref{image60cm}), utilizing a single gas inlet to pressurize a supply channel parallel to the main capillary axis. This supply channel then feeds the main capillary through a set of radial inlets. By carefully tailoring the spacing and diameter of these radial inlets, it is possible to tune the axial density profile. 

\begin{figure}[ht]
\centering
\includegraphics[width=1.0\linewidth]{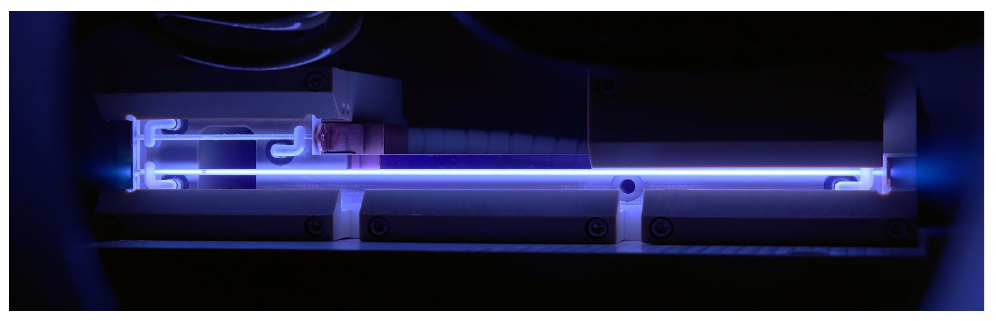}
\caption{High-voltage test discharges for plasma formation inside the 20-cm-long plasma source, where the 5 cm long capillary is used for setup and pre-optimization steps of the longer one.}
\label{Flashcapillary}
\end{figure}

Another solution for making long discharge capillaries to achieve high energies required for the EuPRAXIA project is shown in figure~\ref{Flashcapillary}, which is developed for the FLASHForward experimental facility at DESY to demonstrate plasma acceleration with simultaneous beam-quality preservation and high energy efficiency in a compact plasma stage. It consists of two sections, namely a 5-cm-long capillary made for making precision measurements, as well as the setup and pre-optimization of the source as well as a longer 20 cm section, which is the dedicated source for achieving high energies.

\subsection{\label{sec:discharge-cdw}Capillary Discharge Waveguides}
\subsubsection{\label{ssec:discharge-cdw-bckg}Background}
During the discharge process, in addition to ionization of the neutral gas within the capillary, there is heating of the plasma by the discharge current. 
This heating results in a peaked temperature profile near the axis due to thermal conduction to the capillary walls. 
This temperature gradient results in a density gradient and, under the right conditions, can lead to the formation of a parabolic density profile suitable for guiding high-intensity laser pulses~\cite{Spence2000}.
These sources are then referred to as Capillary Discharge Waveguides~\cite{Spence2003} and they have been critical to the development of GeV level laser plasma accelerators \cite{Leemans2006,Leemans2014,Gonsalves2019}.

\subsubsection{\label{ssec:discharge-cdw-soa}State of the Art}
Capillary discharge waveguides are typically manufactured from sapphire (or sometimes diamond) due to the materials hardness and damage resistance. The diameter of the capillary is on the scale of hundreds of microns to a few millimeters and the length of such capillaries can extend to several tens of centimeters. The stability and reproducibility of these devices has also been investigated~\cite{Turner2021}.

The matched spot size or the fundamental mode of the plasma waveguide is strongly linked to the diameter of the capillary, the operating density and the discharge current characteristics. 
As the desired energy gain from the plasma accelerator increases, the required on-axis plasma density drops. 
This together with restrictions to the capillary diameter required to avoid laser damage leads to the production of waveguides with very large matched spot sizes at low axial densities ($\sim 10^{17}$ cm$^{-3}$) . This can cause a reduction in accelerator efficiency or in the worst case damage to the capillary itself due to leakage of the main beam from the waveguide structure. 
It has been demonstrated that one route to reducing the spot size is to further heat the discharge plasma with an auxiliary nanosecond laser pulse via inverse bremmstrahlung heating. This deepens the plasma channel, reducing the width of the matched spot size and has enabled the production of electrons up to \SI{8}{\giga\electronvolt}~\cite{Gonsalves2019}.
It has also been proposed that the energy gain in such plasma stages can be enhanced via the longitudinal tapering of the plasma channel~\cite{Rittershofer2010}.

To improve the longevity and tunability of capillary discharges waveguides, a regenerative, cryogenically-formed model has been developed which works by freezing a layer of nitrous oxide gas onto the inner wall of a sapphire capillary \cite{Swanson2021}.

\subsection{\label{sec:discharge-apl}Active Plasma Lenses}
\subsubsection{\label{ssec:discharge-apl-bckg}Background}
Active plasma lenses are a promising technology providing compact and symmetric strong focusing magnetic fields by generating large longitudinal currents in thin plasma capillaries. Such devices have been successfully applied to ion beam focusing, and lately to electron beams\cite{Tilborg_PRL_2015,Pompili_APL_2017,Lindstrom2018,sjobak2021strong}.

\subsubsection{\label{ssec:discharge-apl-soa}State of the Art}
\begin{figure}[ht]
\centering
\includegraphics[width=0.8\linewidth]{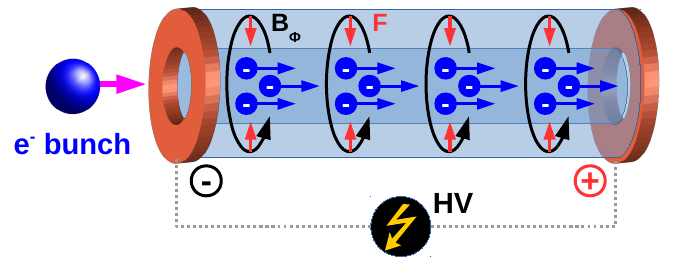}
\includegraphics[width=0.8\linewidth]{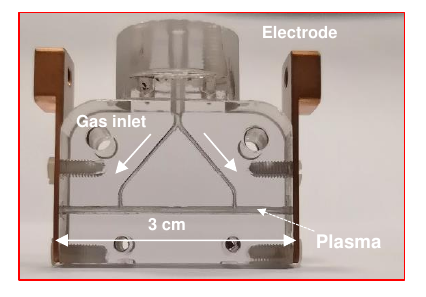}
\caption{Top: Basic concept for explaining the formation of the magnetic field inside an ionized column by electrical discharge. Bottom: Prototype of a gas-filled discharge capillary for use as an active plasma lens. Length and diameter of the plasma channel (3 cm and 1 mm, respectively in this image) are the geometric parameters to be designed according to the focusing capabilities that are desired.}
\label{schemalens}
\end{figure}

Similar to other capillary discharge sources, an APL is typically made up of a few-cm length hollow tube of a few hundred micron to millimeter diameter filled with gas. In the case of light gases, such as $H_2$, typical working pressures of 15-150 mbar are established. As in gas-filled discharge capillaries, a discharge is generated through voltage pulses. Following the gas breakdown, a strong sub-$\mu$s current pulse flows axially establishing the azimuthal magnetic field (Figure~\ref{schemalens}).

APLs are able to provide \si{\kilo\tesla\per\meter} focusing fields, with very low chromatic dependence, if compared to traditional magnetic elements, making them particularly suited for high energy spread beams such as the ones connected to plasma acceleration schemes. 


In general, active plasma lenses are a very useful tool for reducing the cost and size of particle accelerators, since they are able to generate radially symmetric magnetic fields several orders of magnitude larger than conventional quadrupoles and solenoids. 
However, they can only focus charged particle beams without emittance-spoiling aberrations if the current density is constant within the lens.
As such, radial variations in the current density cause significant problems.
The current density is $J_z(r) = \sigma(r) E_z$, where the conductivity $\sigma(r) \propto T_e^{3/2} / \ln(\Lambda)$, where $\Lambda = n_e \lambda_D^3$ and $\lambda_D$ is the Debye length \cite{Tilborg2017}.
Thus the largest factor leading to aberrations is a radially varying electron temperature distribution $T_e(r)$.
It must not vary significantly over the beam size, which is likely to be over order hundreds of micrometers in an APL capture device.
In the steady state, $T_e$ reduces as $r$ increases due to cooling at the wall of the device.
To achieve high focussing gradients, this must not be allowed to happen before the peak of the discharge current is reached.
This challenge was overcome in Ref.~\onlinecite{Lindstrom2018}, where the authors showed that the emittance of the incoming electron bunch could be preserved in an APL by using a heavier gas species, argon in their experiment.
This was possible because the rates of both ion thermal conductivity and electron-ion heat transfer are inversely proportional to ion mass \cite{Bobrova2001}, although the increased effects of scattering off heavier ion species remains to be explored.

\subsection{\label{sec:discharge-challenges}Challenges and Outlook}


General challenges for all capillary discharge plasma sources include scaling the length of these devices to the meter scale and managing the heat load at higher repetition rates, for example via the use of active cooling. Additionally developing improved diagnostics remains a key objective which is made difficult by the closed geometry of the setup (see \ref{sec:diags}). Additionally the longevity of capillaires operated at the \SI{100}{\hertz} level for extended periods has yet to be thoroughly explored. 

In the case of laser guiding structures, the main challenge remains mitigating laser damage and reducing the matched spot size of the guided structure while maintaining a large physical capillary structure at low plasma densities. 

Perhaps the largest challenge for APLs to overcome is the demonstration of emittance preservation for high current electron bunches, whose high densities can drive wakefields in the APL, degrading their emittance.
This is, however, partially mitigated by the fact that the electron bunch is never close to a focus within the device itself.

\section{\label{sec:HOFI}Hydrodynamic Optical-Field Ionised Plasma Channels}
\subsection{\label{ssec:HOFI-bckg}Background}
Starting in 1993, almost ten years before the development of capillary discharge waveguides, Durfee and Milchberg proposed to guide intense laser pulses in an optically-generated plasma channel~\cite{Durfee1993}. The working principle of these plasma channels is very similar to that of capillary discharges, both utilize a radial density profile with a minimum on axis to create a refractive index structure suitable for guiding light \cite{Steinhauer1971}. The difference here, as compared to the capillary discharges, is that the plasma, which expands hydrodynamically to form the plasma channel, is created by a laser pulse instead of by an electric discharge. The main advantage of this all-optical technique is that the plasma does not rely on encapsulation in a solid structure, as in the case of a capillary. Thus, the target can be immune to damage caused by the laser, making it a promising solution for operation at high repetition rates.

In the original scheme, the channel was created by a 100-picosecond laser pulse which was focused using an axicon to produce and heat a centimeter-long plasma filament. The plasma then hydrodynamically expanded, generating a shock at the plasma gas interface. After a few nanoseconds the evolution led to the formation of a plasma waveguide capable of guiding an intense laser pulse over several tens of Rayleigh lengths. In the pioneering experiments, the plasma was generated by collisional ionization seeded by multiphoton ionization. 
Efficiency was greatly improved by using the ignitor-heater scheme in which a femtosecond pulse, the ignitor, ionized the plasma by above-threshold ionization. This plasma was then subsequently collisionally heated by the long pulse, the heater~\cite{Volfbeyn1999,Xiao2004}. This sophistication enabled the first practical application of laser-generated plasma channels, with the efficient guiding of an ultra-intense laser pulse in a high-density krypton gas leading to a dramatic increase in the radiation of plasma-based soft X-ray lasers~\cite{Chou2007,Depresseux2015}.

Due to the use of long laser pulses, and the resulting collisional heating, which is less and less effective as the gas density and Z-number decreases, plasma waveguides generated by lasers were initially used mainly in high Z gases. It, therefore, seemed of little interest for laser-plasma acceleration. However, as early as 1999, the ignitor-heater technique had shown its effectiveness in producing waveguides in low atomic number gases such as hydrogen at electron densities of a few $10^{18}$ cm$^{-3}$. Further studies confirmed this capability while noting that the heater was not necessary and that a guide could be efficiently produced by above-threshold ionization in a low-Z gas using a single few hundred femtosecond laser pulse focused at an intensity of order of a few 10$^{15}$ Wcm$^{-2}$ ~\cite{Lemos2013} or by utilising clustered gases ~\cite{Kumarappan2005}. 

\subsection{\label{ssec:HOFI-soa}State of the Art}
\begin{figure}
	\centering
	\includegraphics[width=0.45\textwidth]{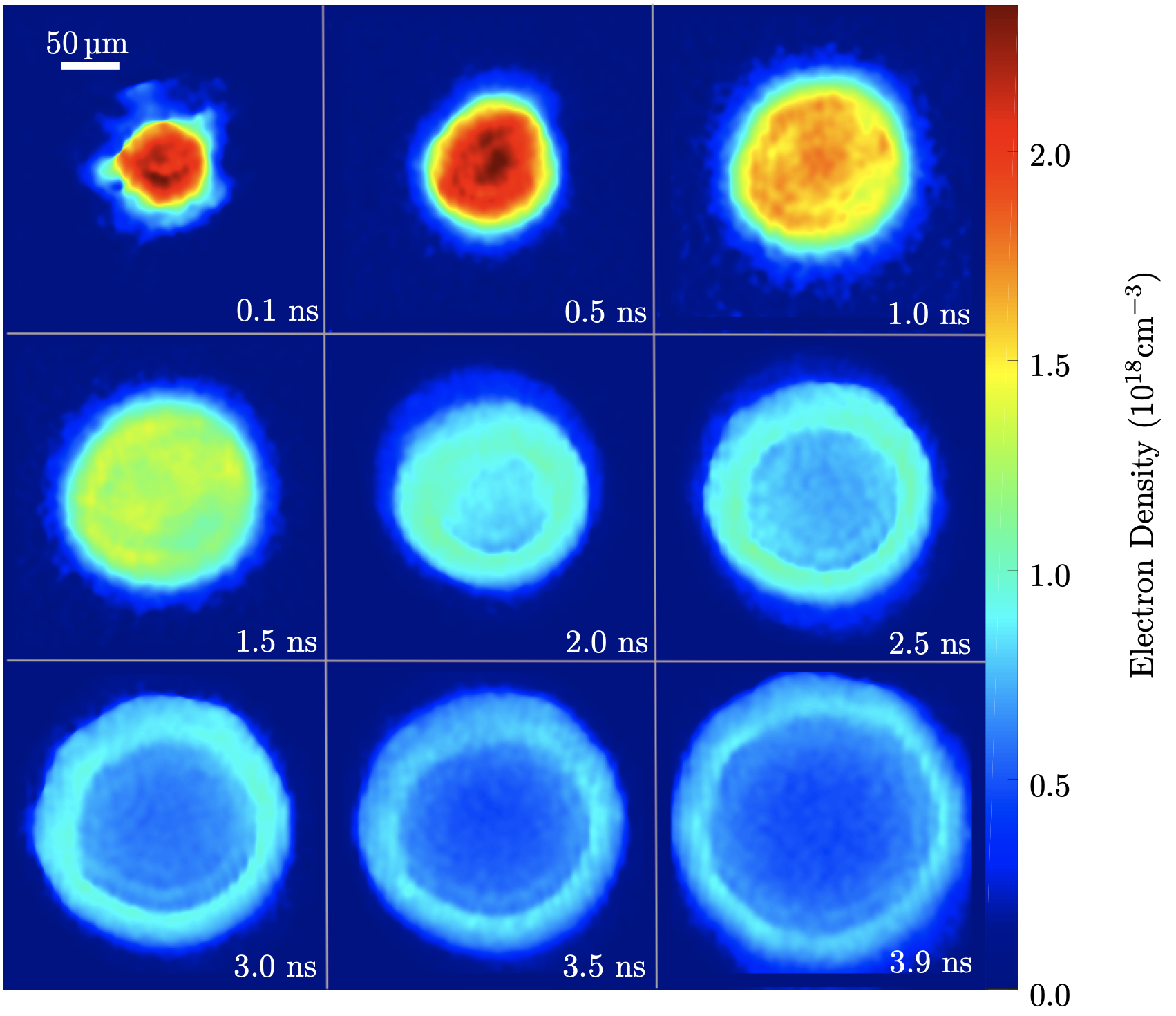}
	\caption{Formation of a HOFI plasma waveguide. Single-shot data showing measured transverse electron density profiles at an array of times after the arrival of a channel-forming pulse focused into 50 mbar of hydrogen gas. Each profile is obtained from analysis of a single interferogram and is shown in a square of side \SI{280}{\micro\meter}. The delay time is indicated for each plot ~\cite{Shalloo2018phd}. }
 \label{fig:HOFI}
\end{figure}

In 2018, it was shown that laser-generated plasma waveguides could be efficiently formed with on-axis densities of $10^{17}$ cm$^{-3}$, an order of magnitude lower than previously demonstrated, enabling their use as waveguides for multi-GeV laser-plasma accelerators ~\cite{Shalloo2018}. The expansion of such as plasma waveguide is shown in fig. \ref{fig:HOFI}. This was made possible through the use of optical field ionization, which heats the plasma independent of density, as opposed to collisional ionization which had been prevalent previously. It was shown that these low-density Hydrodynamic Optical-Field-Ionized (HOFI) plasma waveguides could also be extended in length using an axicon and could guide high-intensity pulses at \SI{5}{\hertz} ~\cite{Shalloo2019}. Further developments led to the demonstration of consistent HOFI waveguide formation at \SI{1}{\kilo\hertz} ~\cite{Alejo2022}.

As the plasma expands radially outwards, it drives a neutral density shock in addition to the electron density shock. This neutral collar of gas surrounding the expanding plasma waveguide typically has a density larger than the ambient gas and a significantly higher density than the axial electron density. Thus, when a high-intensity femtosecond pulse is guided within the channel, its leading edge can field-ionize this collar of gas and thus significantly increase the electron density peak in the walls of the waveguide. This improves the guiding characteristics including stronger confinement leading to low-loss meter scale plasma channels ~\cite{Shalloo2018phd,Morozov2018,Picksley2020,Feder2020}. This collar of gas can also be pre-ionised with an auxiliary tailored pulse ~\cite{Miao2020}.

Generating HOFI channels requires producing a quasi-Bessel beam, i.e. a laser beam with an extended depth of focus, which is traditionally done using an axicon lens. It can, nevertheless, be advantageous to use diffractive optical elements (DOE), which can be more compact and precise. Diffractive axicons have already been successfully substituted for their refractive equivalent~\cite{Feder2020}, but the possibilities offered by DOEs are much more extensive. For example, axilenses~\cite{Davidson1991} could make the use of laser energy more efficient and allow for shaping of the longitudinal intensity profile of the quasi-Bessel beam, thus offering a new possibility for optimizing guiding. The axiparabola, which is the reflective counterpart of axilenses, can offer the same flexibility while being achromatic and having better flux resistance~\cite{Smartsev2019,Oubrerie2022}.

The first demonstration of multi-GeV electron beams from an hydrodynamic plasma waveguide generated via optical field ionization was achieved by Miao \emph{et al.} \cite{Miao2022}. A \SI{300}{\tera\watt} laser was guided with a \SI{20}{\centi\meter} waveguide producing \SI{5}{\giga\electronvolt} beams with $\sim$ 15 \% relative energy spread. This has recently been extended up to the \SI{10}{\giga\electronvolt} level \cite{Picksley2024}.

While experimental work on HOFI channels has advanced rapidly in the last few years, the simulation of such plasma sources has progressed more slowly. This is due to the complexity of accurately modeling a variety of process on a range of different timescales; ionization and heating on femtosecond timescales, thermalization on picosecond timescales and expansion and channel formation on nanosecond timescales. Recently significant progress has been made in developing start-to-end simulation frameworks benchmarked against experimental data ~\cite{Mewes2023,Miao2024} although an open source codebase is still lacking. 

As the generation of the plasma guide is based on optical field ionization, the generation efficiency does not depend on the plasma density. This, along with the absence of any need to have a structure around the plasma, offers significant flexibility in shaping the longitudinal density profile. These properties can be used to control the trapping of electrons through density-transition injection\cite{Bulanov1998, Faure2010,Suk2001a, Schmid2010}. Oubrerie \emph{et al.} achieved the first demonstration of controlled injection in a HOFI plasma waveguide by placing an obstacle above a supersonic jet to produce a hydrodynamic shock associated with a transition density in the neutral gas profile and then forming a HOFI waveguide in this structure~\cite{Oubrerie2022b}. The electron beams accelerated in this work, with a 60 TW laser, have a peak energy of about 1 GeV, with relative FWHM energy spreads as low as 2\%. The control of the injection also improves conversion efficiency, with several percent of the laser energy transferred to GeV electrons. Alternatively, Picksley \emph{et al.}  obtained a sharp down-ramp by using a truncated channel, achieved by moving the longitudinal position of the start of the HOFI plasma channel within the neutral gas.  This scheme produces electron bunches with an energy of up to 1.2 GeV and an RMS energy spread of 4.5\% with 120 TW laser pulses~\cite{Picksley2023}. A further scheme has been proposed which leverages hydrodynamic plasma structuring to tailor the waveguide properties longitudinally and thus control injection\cite{Shalloo2024}.

\subsection{\label{ssec:HOF-I-challenges}Challenges}
HOFI waveguides are a promising technology for the generation of high-quality GeV-class electron beams from laser-plasma accelerators. However, for deployment at a EuPRAXIA laser-driven facility, several key challenges must be tackled including; increased laser stability, improved plasma density tailoring, accelerator vacuum integration and further improvements in electron beam quality. In particular, with regard to vacuum integration for a multi-GeV HOFI stage, it is important to consider the delivery mechanism for gas to the interaction region and whether this flow rate can be sustained during high repetition rate operation. The gas delivery / confinement method should also avoid interference with the plasma channel forming laser pulse. Development of such a gas delivery module suitable for EuPRAXIA parameters is an open challenge. 

\section{\label{sec:PM}Plasma Mirrors}
\subsection{\label{ssec:PM-bckg}Background}

Plasma mirrors use rapid ionization by high-intensity lasers to form a reflective plasma on the surface of a material. They have been used extensively for contrast enhancement of ultra-short laser pulses over the past two decades \cite{Dromey2004RSI}. In this case, the initial substrate has an anti-reflection coating to reduce pre-pulses and amplified spontaneous emission by more than two orders of magnitude (see, e.g., Ref. \cite{Rodel2011APB}). The reflected pulses have demonstrated very high spatial quality and operation for pulse lengths from tens of femtoseconds up to the picosecond regime.

In laser wakefield acceleration, it is desirable to use plasma mirrors to couple in and out the high intensity laser pulses before and after accelerating stages in order to minimise inter-stage distances. They have also been used purely to remove the laser pulse after a single stage to enable applications using the betatron x-ray beam \cite{Cole2018PNAS}, or for secondary particle production (e.g. positrons) using the generated electron beam \cite{Streeter2024SR}.

The laser-driven EuPRAXIA beamline will need to remove the laser from the axis after the plasma accelerator to protect components immediately downstream, such as an active plasma lens or quadrupole magnets. Additionally, it is advantageous to use the reflected beam as an exit mode diagnostic to characterize and monitor the performance of the accelerator. Because of the long focal length of the drive laser focusing optic, a second plasma mirror is useful to direct the diagnostic beam anti-parallel to the incoming laser to maintain a small width footprint as is usual for accelerator beamlines.

To be suitable for high repetition rate operation, plasma mirrors used for EuPRAXIA will need to be formed on continuously refreshed targets rather than the fused silica substrates commonly used for contrast enhancement in solid target interactions. Two technologies are capable of this: tape drives and liquid sheets. Both types of targets have been studied and used on the high energy (multi-GeV) laser wakefield electron beamline at the BELLA centre to reject the laser pulses in order to protect their active plasma lens \cite{Zingale2021PRAB}. 

\subsection{State of the Art}\label{ssec:PM_stateoftheart}
Tape drives are a well-established technology and are constructed by spooling a tape across the interaction region at a suitable speed \cite{Shaw2016POP,Noaman2017PRAB,Condamine2021RSI,Robins2022CLF,Xu2023HPLSE,Zeraouli2023RSI,Ehret2024PPCF}. The challenge for laser-plasma interactions is the precision required for repeatability in position and angle, as well as the surface quality and flatness. The tape is passed over two pins and forms a flat surface by tensioning the tape with the spool motors. An additional support plate with an aperture at the interaction point can be added to hold the tape. In this way, a longitudinal position stability of better than 5 microns has been achieved \cite{Xu2023HPLSE,Robins2022CLF,Zeraouli2023RSI}. A curved surface can also be produced by bending the support plate \cite{Zeraouli2023RSI} and this could be useful in the design of exit mode diagnostics.  

Characterization of tape drives \cite{Shaw2016POP, Robins2022CLF} has shown that the reflected beam quality and pointing (of order 1 mrad) is acceptable for use in an exit mode diagnostic, with no significant degradation in focal spot quality after reflection. Various plastic materials have been tested with the best surface quality consistently measured with VHS tape (surface roughness < \SI{5}{\nano\metre} Ra). Because of the difficulty in sourcing VHS tape and the presence of impurities, a more practical choice is polyimide tape (surface roughness < \SI{10}{\nano\metre} Ra), which is readily available at low cost. The surface must be flat (or consistently curved) over the area which the drive beam irradiates. This is easily achieved close to the laser focus but is more challenging over millimetre-scale areas, so the positioning of the plasma mirror must be carefully considered.

Several groups have an active program of developing tape drives \cite{Robins2022CLF, Zeraouli2023RSI, Ehret2024PPCF, Condamine2021RSI}, predominantly for short-focus solid target interaction experiments and high repetition rate plasma mirrors. These have a more stringent requirement on longitudinal positioning and refection quality than the laser-wakefield application so these systems would be suitable for EuPRAXIA.

In terms of technological readiness, tape drives are a straightforward solution to implement but have several major drawbacks. The first is simply the amount of tape required for continuous operation at the \SI{100}{\hertz} level which is of order 10 km per hour. Secondly, the standard \SI{12.5}{\micro\metre} thick plastic tape would produce significant amounts of debris (10s grams per day). Finally, the thickness of the tape causes a significant increase in beam emittance \cite{Zingale2021PRAB}.

At high (\qty{>10}{Hz}) repetition-rates, continuous operation of tape targets become a critical issue.
In this regime, liquid sheet targets offer a practical alternative to tape targets, addressing the material supply and coating problems simultaneously.
Three approaches to forming suitable liquid sheets have been demonstrated so far. These include; the use of liquid crystal targets, e.g. \cite{Schumacher2017JInst,Zingale2021PRAB}, the use of colliding liquid streams, e.g. \cite{Morrison2018NJP,George2019HPLSE,Kim2023NC,Fule2024HPLSE} and the use of shaped nozzles \cite{Treffert2022APL,Treffert2022POP}.

As mentioned above, the presence of a plasma mirror in the electron beam path can induce a non-negligible degradation of the electron beam emittance through scattering and magnetic field generation \cite{Raj2020PRR}.  Scattering can be minimized by reducing the thickness of the plasma mirror material. In this context liquid sheet targets can be advantageous. For example Ref. \cite{Zingale2021PRAB} experimentally showed that an ultrathin (20nm) liquid crystal plasma mirror can induce negligible emittance degradation. 

Such targets have demonstrated high surface quality \cite{Kim2023NC}, compatibility with high-intensity laser-plasma interactions \cite{Treffert2022APL} and sheet thicknesses of \qty{<100}{nm} \cite{Treffert2022POP,Fule2024HPLSE}.
However, only the colliding liquid streams and shaped nozzles methods have demonstrated kHz compatible operation \cite{George2019HPLSE}.
Increasing the repetition rate of liquid crystal targets to 100 Hz would be a challenge.
Liquid sheets present additional challenges, as the liquid flow in vacuum must be controlled while the vapour must be removed to maintain vacuum pressure.
Furthermore, the sheets are typically limited to mm transverse sizes, so they can not be placed arbitrarily far from the plasma exit and still capture the full transverse extent of the diverging laser pulse. 


\subsection{Challenges}\label{ssec:PM_challenges}


There are several challenges associated with the use of plasma mirrors. These include, as mentioned above; scalability to high-repetition rates, vacuum integration considerations, and deleterious effects on the transmitted electron beams. 

\begin{figure}
	\centering
	\includegraphics[width=0.45\textwidth]{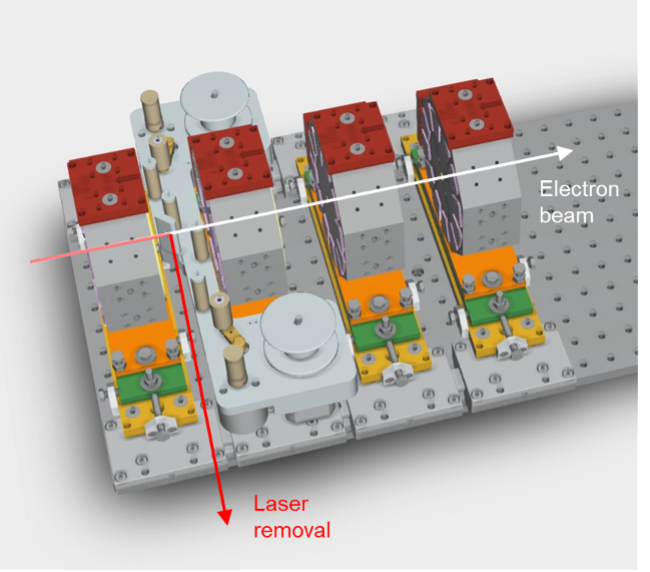}
	\caption{Tape drive plasma mirror installed between two quadruple magnets. The magnet widths and gaps between them are both 50mm.}
 \label{fig:PM_system}
\end{figure}

Another key challenge is optimizing the geometry of a plasma mirror system, in particular the distance from the exit of the plasma accelerator. The strength of the magnetic field generated inside the plasma mirror increases with laser intensity. If the plasma mirror is placed too close to the exit of the plasma module, this will cause a significant increase in divergence. This can be simulated using particle-in-cell codes and has been experimentally studied by \cite{Raj2020PRR}.
The bandwidth of the electron spectrum can be also be increased by plasma mirrors, due to inverse-Compton scattering in the reflected laser pulse \cite{Streeter2020PRAB}.
If the plasma is created by the drive pulse itself, the distance of the plasma mirror from the exit of accelerator must be chosen to optimize the incident laser intensity. 
This is non-trivial because the temporal and spatial profile of the driving laser pulse will be modified by the interaction within the plasma accelerator. 
This can be avoided by triggering the plasma mirror with a second laser pulse \cite{Indorf2022PPCF} which could produce consistent plasma mirror conditions, reduce drive pulse energy transmission, and reduce debris. However, the second pulse would need to be precisely synchronized and focused to optimal intensities meaning the plasma mirror needs to be close to the exit of the accelerator. 
Additionally, the positioning requirements of other components, such as quadrupoles or APLs could conflict with the plasma mirror positioning requirements, constraining the problem. This can be mitigated by integration of the plasma mirror within other post interaction components as shown in fig. \ref{fig:PM_system}.


\section{\label{sec:materials}Material Robustness and Longevity}

The development of robust plasma accelerator components enabling daily operation at a high repetition rate ($\geq 10\,$ Hz) with a beam up time availability rate compatible with a user facility is an outstanding challenge for the materials used for the gas confinement (cell\cite{Drobniak2023a}, discharge\cite{Gonsalves2016}) or gas injection (nozzles \cite{Prencipe2017, Tomkus2018, Rovige2020}). The fragility of plasma components lies in the gas container or the gas delivery system. The reduced apertures (flow rate scales with $d^4$, $d$ being the aperture diameter) for the gas cells or, to a lesser extent, the low height of the laser focus from the jet can limit their lifetime. 
It has been shown that shaping the density profile is important for accelerating high-quality electron beams. Shaping of the neutral gas flow is made possible by optimizing the apertures or channels for the cells or by micro-structuring the nozzles for the jets\cite{Tomkus2019,Tomkus2024}.  Geometric alterations, by thermal load or ablation,  to the apertures or nozzles will, over time, change the density distribution and hence the properties of the components. 
Several phenomena are involved in the stability and wear of components. For all components,  the materials surrounding the plasma absorb the power radiated by the plasma.  Secondly, in the case of laser-driven components, the scattering and diffraction of the laser driver can cause some of the laser power to dissipate into the adjacent walls or nozzles\cite{Tomkus2020, Prencipe2017}.
To date, plasma systems have been tested and used for electron beam production by wakefield acceleration with an average driver power of around ten watts for laser-driven systems due to the stability limitations of class $100$~TW laser drivers when operated at rates above Hz.  Demonstrations of stable operation of laser-plasma electron sources with more modest mJ-class energy systems operating at kHz are encouraging. For laser-driver systems, the possibility of testing at higher average power is imminent with projects such as LAPLACE\cite{Laplace-HC} and KALDERA\cite{Kaldera}. For beam-driven systems, working with trains of bunches at a modest rate may limit the thermal effects. More advanced studies on material and power dissipation within plasma components are required to operate in the kW range. Here we briefly discuss laser ablation issue for laser-driven components and wall erosion in the case of discharges. Additionally, we explore thermal management and advanced manufacturing technqiues.

\subsection{\label{sec:materials-lidt}Laser-Induced Damage}

The case of laser-driven plasma components is particularly challenging as the laser energy distribution in the transverse plane at distances far beyond the waist size can exhibit fluences $>1$ J.cm$^{-2}$) exceeding the ablation thresholds of metals \cite{Gamaly2002} and reaching the limit for dielectrics\cite{Nieto2015}. The pulse duration of the laser drivers is in the femtosecond range, leading to a non-thermal ablation regime with a laser-solid type interaction where the absorption of the material is intra/inter-band in the skin thickness of the solid\cite{Gamaly2002}. The development of femtosecond laser machining with athermal ablation and high efficiency, when GHz burst pulses are used \cite{Kerse2016}, drives numerous studies on material ablation properties in single and multiple pulses of interest in material for laser-driven plasma components. For example, high-Z materials like tungsten or ceramics exhibit high ablation threshold \cite{Schroeder2016,Hazzan2021}. Also, ablation can be saturated at very high-intensity laser \cite{Zhang2011,Zheng2024}. 
In plasma accelerator experiments using gas cells with sub-mm metal apertures, it is quite common, depending on laser-pointing stability and focusing geometry, for the apertures to enlarge after several thousand high-intensity shots.  
Overall optimization could improve the robustness of plasma components to laser radiation by linking geometric constraints to the ablation threshold or by maximizing laser absorption or reflection on the most exposed parts. Dedicated studies or more documented tests are needed to achieve a satisfactory technical readiness level.

\subsection{\label{sec:materials-discharge}Damage in Discharge Systems}

For gas-filled capillary discharge made of sapphire or fused silica, the wall erosion is negligible on the scale of an operation day\cite{Gonsalves2016}. For the EuPRAXIA parameters for plasma wakefield acceleration, the power deposited by the pulsed discharge per unit of capillary length is not considered an issue for the envisaged repetition rates if appropriate choices of cooling and materials are made\cite{Sasorov2024}. Nevertheless, if capillary discharge systems are considered for guiding $100$~TW class laser-driver at a high repetition rate, a significant fraction of the laser driver energy can be transferred to the capillary wall \cite{Gonsalves2019}. 

\subsection{\label{sec:materials-thermal}Thermal Management}

For plasma components aiming to operate at $1\,$kHz repetition rate, the total dissipated power can reach several hundreds of Watts to the wall of the capillary, cells, or differential pumping chamber and nozzle for the gas jet. Operating plasma systems reported here, all operate in the tens Watt regime. The availability of high-average power drivers is crucial for developing reliable plasma components for EuPRAXIA. 

\subsection{\label{sec:materials-amt}Advanced Manufacturing Techniques}

The critical parts of plasma components; apertures, nozzles and cells require precision and sometimes complex shapes in small dimensions. Femtosecond Laser-assisted Selective Etching is used to produce sub-millimetre nozzles\cite{Tomkus2018,Tomkus2024} (see sec.\ref{ssec:gastarget-chall}) and femtosecond laser engraving and cutting using Bessel beam with a few micrometre precision of channels for cells\cite{Drobniak2023b}. Additive manufacturing was first used to manufacture nozzles using polymers\cite{Dopp2016,Andrianaki2023} and is now tested using metals. In recent years, additive manufacturing has seen spectacular growth, particularly in manufacturing precision parts from tungsten\cite{Howard2024, Hazzan2021}, glass\cite{Xin2023} and ceramics\cite{Dadkhah2023}. These advances open up new possibilities in designing and developing plasma components.

\section{\label{sec:integration}Plasma Component Integration}

In the future, the question of integration arises from the point of view of vacuum, alignment, and coupling to the beamline to implement plasma sources or components in a machine for user access. The plasma sources and systems discussed in sections \ref{sec:discharge},\ref{sec:gastarget}, and \ref{sec:HOFI} are generally integrated into one of two ways: (i) the laser focusing elements, the plasma source and the diagnostic or electron beam capture section are integrated into a large high vacuum chamber with a typical volume of $\sim 1\,$m$^3$; (ii) the plasma source or accelerating stage component is integrated into a smaller vacuum chamber and forms part of the accelerator beamline. These two approaches lead to various vacuum and alignment constraints. 

\subsection{\label{sec:integration-vac}Vacuum Integration}

Operating particle accelerators under vacuum is essential for minimizing particle interactions with atoms and molecules, preserving electron beam quality, protecting accelerator components, and ensuring accurate beam transport \cite{Dabin2002, Malyshev2019}. A high vacuum level is a fundamental requirement for precise control and high performance in modern electron beam linac. In most of the laser-driven plasma electron sources, due to the compactness and the interaction chamber experimental setup, the vacuum level ranges from $10^{-6}$ to $10^{-3}\,$ mbar.

As mentioned in the previous sections, plasma systems are based on gas injection under vacuum, either in pulsed or continuous mode.  The integration of plasma systems requires a differential pumping system between the region where the gas is injected and the rest of the beamline\cite{Luo2021, Storey2024}. Differential pumping systems have been used for years to implement windowless gas targets in accelerator lines for nuclear physics\cite{Sagara1996, Luo2021}, and low-density plasma systems have been used for beam-driven plasma wakefield in various proof of principle experiments\cite{Vafaei2013, Adli2018, Lindstrom2024} allowing insertion in ultra-high vacuum beamline. 
Differential pumping setups are now adapted for gas jets\cite {Monzac2024a} and gas cell laser-plasma sources\cite{Messner2020, Drobniak2023b}.  The reduction in conductance is achieved by using small orifices, narrow tubes or openings between chambers to restrict the gas flow, thus maintaining the pressure difference.  Vacuum pumps are used at each stage to obtain the desired pressure gradient. The choice of the pump speed depends on the vacuum level required and the gas type to be pumped \cite{Schurmann2013}.  Depending on the arrangements and the gas flows considered, 4 to 6 orders of magnitude pressure drops can be obtained over a few tens of centimetres\cite{Messner2020, Monzac2024b}.

\subsection{\label{sec:integration-align}Alignment and beamline integration}

The scope of the present section is not to review the state-of-the-art mechanical alignment of particle accelerators but to highlight the particular aspects of aligning plasma components in an accelerator.  High-precision mechanical alignment of particle accelerator~\cite{Martin2010} involves a combination of high-precision measurement tools, an automated system, and continuous environment monitoring to achieve and maintain alignment accuracy necessary for optimal performance~\cite{Gonzalez2022}. High gradient plasma components reduce the length of accelerating sections,  leading to a typical full beamline length for EuPRAXIA plasma-based FEL of $\sim 145 \,$m~\cite{Ferrario2018}. As an example beam-driven site \textsc{EUPRAXIA@SPARC\_LAB} layout is composed by a $55\,$m length linac, $\sim 4\,$m plasma module, $\sim 40\,$m undulators and finally $50\,$m long FEL user beamline. The laser-driven EUPRAXIA FEL is expected to be slightly shorter depending on the layout considered \cite{Assmann2020}.         

Typical working dimensions of plasma components range from hundreds of micrometres to millimetres in the transverse direction and from millimetres to tens of centimetres in length. Aligning the target in the driver, electron, or laser beam requires 3 axes of translation extending up to 5 axes if the two rotations around the transverse axes are added. Technical solutions for moving targets and plasma components based on stepper motor drivers with good resistance to electromagnetic radiation and noise are generally preferred. 


The combination of a reference network, a laser tracker, and a coordinate measuring machine~\cite{Martin2010, Marsh2018} can allow for high precision mechanical positioning ($\pm 100\,\mu$m) of plasma components. In the case of the laser-driven source, the capture section design often involves a high-gradient quadrupole doublet or triplet~\cite{Assmann2020, Winkler2019} or an active plasma lens~\cite{Ferranpousa2019} close to the accelerator. This requires a high precision alignment and coincidence of mechanical, magnetic and optical reference axes. The reference of the transverse and longitudinal laser focal spot positions in the mechanical reference network can be transferred using a high-magnification imaging system combined with mechanically scanned fiducials or calibrated tips. A low energy alignment laser can be combined with a Bessel beam~\cite{Parks2021}, or Structured laser beam \cite{Polak2022} for high precision optical transverse component alignment all along the beamline with an appropriate imaging system and apertures~\cite{Gayde2023, Niewiem2023}. 
One should notice that for a laser-driven electron source, full in-vacuum integration of the gas target and first optics of the capture section~\cite{Labat2023} on motorized translation stages, relaxes the alignment tolerances and allows the use of alternative beam alignment techniques, like beam pointing alignment compensation~\cite{Andre2018}. 
Alignment and beamline integration constraints can be mitigated by a more compact plasma acceleration stage with two integrated short focal length active plasma lenses for coupling to the plasma acceleration stage~\cite{Pompili2024}. 

The vacuum integration and alignment of plasma systems in future plasma-based accelerators are key to providing high-performance and high-quality beams. Recent progress in this area in various plasma system configurations is very encouraging for the implementation of a robust solution for EuPRAXIA.

\section{\label{sec:diags}Plasma Diagnostics and Tools}
\subsection{\label{sec:diags-laser}Laser-Based Diagnostics}
There exists a wide variety of laser-based diagnostics for investigating plasma properties. Some important examples for investigating the initial plasma conditions in an accelerator module include, for example; interferometry, shadowgraphy, schlieren imaging and laser-induced fluorescence. 
They are used to measure/monitor (quantitatively and qualitatively) the particle number density ($n_g$ for neutral gas and $n_e$ for free-electrons in a plasma) and its fluctuations \cite{Brandi2019,Harilal2022}.  The diagnostics presented here focus on measuring the bulk plasma properties, however there are a variety of techniques for probing the plasma accelerator structure. An excellent review of such techniques can be found in Ref. \cite{Downer2018}.

\subsubsection{\label{ssec:diags-interferometry}Interferometry}
For a plasma density well below the critical density ($n_e \ll n_c = m_e \epsilon_0 \omega^2/e^2$) and for weakly interacting gasses, the medium refractivity $(\eta-1)$, where $\eta$ is the refractive index, is with excellent approximation, proportional to the particle number density. Therefore the measurement of the phase shift of an electromagnetic wave traveling through the plasma source allows the determination of the line-integrated particle number density.

For plasma accelerator modules, this phase shift is often extracted via interferometry and the electromagnetic waves used to probe the plasma are typically femtosecond to picosecond laser pulses, termed probe pulses, sychronised to the plasma forming laser/discharge. Interferometric measurements are often made with the probe pulse traveling transverse to (transverse interferometry), or in line with (longitudinal interferometry), the accelerator axis.

A variety of interferometer designs can be used to diagnose the plasma density \cite{Brandi2019}.
The two interfering pulses are often obtained by splitting a single laser beam. This can happen prior to passage through the sample, i.e. the beams travel along two different paths; one beam probing the sample and one acting as a reference. Alternatively, the two interfering pulses can be separated after the sample provided only a portion of beam passed through the sample. This can be achieved by using an interferometer in a folded wavefront configuration. The latter cases being less sensitive to environmental conditions, more compact, and therefore in general more commonly adopted in research laboratories. Some common folded wavefront interferometer configurations utilised for plasma density measurements include; Mach-Zehnder, Michelson, Nomarsky and Quadriwave lateral shearing interferometers. For non-folded wavefront variations Mach-Zehnder interferometers are often used.  

The configurations above measure a projected 2D spatial distribution of a 3D plasma density structure. In the case of transverse interferometry, this projection is often assumed to be symmetric about the accelerator axis. In this case, the 3D density map may be retrieved with the well known Abel inversion technique \cite{Abel1826}. For non-symmetric samples a tomographic approach can be implemented with multiple acquisitions. Modified Abel inversion techniques have also been investigated for simple asymmetries \cite{Shalloo2019}.

Adding a second wavelength to this interferometric approach can be used to determine the density of neutrals within the plasma, in addition to the free electrons \cite{Alcock1966,Feder2020}. The specific neutral gaseous media used in plasma acceleration modules can be composed of a single species (e.g., nitrogen, hydrogen), or mixtures (e.g., helium with nitrogen doping at the percent level) which are widely adopted to achieve ionization injection. For such kinds of gas mixtures the total phase shift of the probe pulse passing through the medium is the sum of the phase shifts from the free electrons and from each individual ionic species.

An important interferometric technique for line-integrated plasma density measurements is second-harmonic dispersion interferometry (SHDI)\cite{Brandi2007,Brandi2016,Tilborg2018,Tilborg2019,Garland2021}. SHDI is based on a common-path design and it is already widely adopted in magnetic confined fusion machines due to its stability and high phase sensitivity also when operated in harsh environments \cite{Hopf1980, Drachev1993, Brandi2009, Brandi2020,Brandi2020a}. In brief, a probe pulse is frequency doubled to create a pair of probe pulses, a fundamental pulse and its second-harmonic, which then co-propagate through the sample. 
After the sample, the fundamental pulse is frequency doubled again, and finally interference takes place between the two second-harmonic beams. 
The measured phase shift is proportional to the medium dispersion, i.e., the difference of the refractive index between the second harmonic and fundamental wavelengths, $\Delta \eta = \eta_{\lambda/2}- \eta_{\lambda}$. 
These phase shifts can be probed in the time-domain \cite{Brandi2007,Brandi2016} or in the spectral domain \cite{Tilborg2018,Tilborg2019,Garland2021}.

\subsubsection{\label{ssec:diags-shadowgraphy}Shadowgraphy and Schlieren Imaging}
Other laser-based diagnostics used to monitor plasma acceleration modules are shadowgraphy and schlieren imaging. Shadowgraphy \cite{Schoebel2022}, as the name suggests, acquires the "shadow" of the medium when illuminated with a laser by monitoring variations in the transverse intensity profile of the laser. Since the medium are usually not absorbing at the probing wavelength, the intensity pattern generated by the sample is due to scattering by in-homogeneity in the density; shadowgraphy is sensitive to the second derivative of the refractive index distribution and therefore to the derivative the particle number density gradient, and it is useful to "image" the medium in-homogeneity and highlight the medium border with respect to the surrounding medium, e.g., free-electron density in a laser-generated plasma. In schlieren imaging a sharp blade is used to block the undisturbed part of the probing laser beam after passing through the sample and before reaching the detection camera. Schlieren imaging is sensitive to the gradient of the refractive index distribution in the medium and it is suitable to monitor the density instability due to hydrodynamic turbulence in the particle flow \cite{Raclavsky2024}. Both shadowgraphy and schlieren imaging can be implemented in parallel with an interferometer by splitting the probing laser beam after passing through the sample, and therefore are very useful auxiliary diagnostics to monitor the sample density fluctuation.

\subsubsection{\label{ssec:diags-lif}Laser-Induced Flourescence}
Laser-induced fluorescence has been recently adopted as an alternative approach to accurate characterization of plasma accelerator modules\cite{Fan2020}. In this diagnostics a laser beam focused to a line with a cylindrical lens is used to excite the molecules of a plane inside the medium and then the fluorescence light emitted by the excited molecules is detected giving information on the local particle density map directly. Laser-induced fluorescent is a very sensitive and background-free technique, and can be used to measure particle density also for non-symmetric samples; the excitation beam wavelength and line-width has to match the absorption resonance of the molecule in the sample.\\
    
\subsection{\label{sec:diags-pressurebroadening}Pressure-Induced Spectral Broadening}
Passive diagnostics based on the properties of light emitted from the plasma are attractive as they are simple to implement and maintain.
This is particularly beneficial if the plasma device resides in an area with highly restricted access such as an accelerator tunnel.
Both the density and temperature of the plasma can be inferred from the pressure-induced spectral broadening of recombination light emitted by the decaying plasma \cite{Griem2012, Harilal2022, Garland2021}. 
The broadening occurs because the ionic potential is altered by the electric field of a nearby electron or ion.
Hydrogen gas is commonly used due to the relative simplicity of interpretation of the broadening of the H$_\alpha$ line \cite{Harilal2022,Parigger2018}.
If other gas species are to be studied a few per-cent hydrogen dopant is added as a `tracer', although care must be taken during analysis in case other emission lines overlap with the hydrogen lines.
The broadening of a single line is mostly a function of plasma density, while the relative amplitudes of multiple lines can be compared to calculate the plasma temperature \cite{Aragon2008, Kunze2009}.
With this diagnostic, ease of implementation is traded off against the difficulty of data interpretation, as the relationship between the line width and the plasma density is model-dependent \cite{Mijatovic2020, Gigosos1996}. 

For a minimally invasive diagnostic, light can be transported in a fibre to an easily accessible spectrometer. 
Longitudinal resolution may be obtained by performing a scan of the plasma source position with respect to the collection optics \cite{Garland2021}.
Alternatively, the plasma can be imaged on to the slit of an imaging spectrometer to obtain longitudinal resolution in a single image.
Due to the large ratio of length to width of most discharge- or laser-created plasmas, an imaging system based on cylindrical lenses can be considered.
The measured line width is usually approximated as a Voigt function, where the instrument function (a gaussian) is convolved with a Lorentzian, whose width varies as a function of plasma density.
Good agreement has previously been found between the density inferred from line broadening and that measured by interferometric methods \cite{Garland2021}, following the modelling in Ref.~\cite{Gigosos1996}.
In typical setups \cite{Garland2021}, the minimum resolvable plasma density using this technique is a few $\times 10^{15}$\,cm$^{-3}$.

\subsection{\label{sec:tools-sim}Simulation tools}


\subsubsection{\label{ssec:tools-cdf}Computational Fluid Dynamics}

As discussed in section \ref{sec:gastarget}, neutral gas targets are critical for plasma components and systems and as such, great importance is placed on the modeling of these sources using for examaple computational fluid dynamics (CFD). The geometry of such targets consists of a pressurized gas inlet, the shape of the target body in which the gas propagates, and a gas outlet in the form of a large vacuum chamber surrounding the target. Based on the rarefaction of the flow, the gas flow dynamics in the target should be modelled by either the statistical mechanics or the continuum mechanics formulation. This is determined by the so-called Knudsen number $\kappa_n= \lambda /L$, defined as the ratio of the molecular mean free path length $\lambda$ to a representative length scale $L$.
When the Knudsen number is small ($\kappa_n < 0.01$), non-equilibrium effects are insignificant, and the standard Navier-Stokes (N-S) equations can accurately predict the gas flow behaviour. As Kn increases ($0.01 < \kappa n < 0.1$), non-equilibrium regions appear near surfaces as the molecule-surface interaction frequency is reduced - the most recognizable effect is velocity slip and temperature jump. The N-S equations can still be used effectively, but the slip and jump boundary conditions must be implemented. However, once the $\kappa_n$ increases into the transition continuum ($0.1 < \kappa_n < 10$) and free-molecular regimes ($\kappa_n> 10$), the N-S equations cannot predict the gas behaviour. The Direct Simulation Monte Carlo (DSMC) method can be used for these purposes.
The CFD approach, which solves Navier-Stokes equations, is sufficient for 2D and 3D gas flow simulations through most gas targets, including a small part of the vacuum chamber close to the target exhaust \cite{Lorenz2019}. CFD codes currently used are OpenFOAM \cite{OpenFOAMpaper,OpenFOAMwebpage}, ANSYS Fluent \cite{ANSYSwebpage}, COMSOL Multiphysics \cite{COMSOLwebpage}. Although incompressible flow models or 2D geometries have been used in the literature \cite{Guillaume2015, Rovige2021}, more accurate results in the regimes of interest for plasma accelerators are obtained using compressible flow models \cite{Audet2018, Aniculaesei2018, Drobniak2023phd, Drobniak2023phd, Messner2020, Tomkus2024}. This is particularly true for the simulation of targets with sharp density transitions, e.g. gas jets for shock injection \cite{Guillaume2015, Rovige2021}. The density distribution obtained through a CFD simulation can be easily used in particle-in-cell (PIC) codes for the next stage of numerical modeling involving kinetic effects in plasmas, as discussed in the next section.

\subsubsection{\label{ssec:tools-pic}Particle-in-Cell Codes}
The beam dynamics of the accelerated electron bunches in laser-driven and beam-driven plasma accelerators requires modeling at the kinetic scale. Such kinetic modeling is often infeasible for multi-dimensional realistic cases using Vlasov codes, which directly solve Vlasov equation. This is one of the reason why the Particle in Cell (PIC) method has become the most common technique to simulate electron wakefield acceleration. Other reasons for its widespread use include the simplicity of its basic formulation, its versatility and its scalability through high performance computing techniques. 

PIC codes used for plasma accelerators compute an approximate, self-consistent solution of the coupled Vlasov equation and Maxwell's equations. To tackle this task, they discretize the space where the electromagnetic fields, current and charge densities are defined with a numerical grid, and sample the plasma distribution function in the 6D coordinate-momentum space with macro-particles. 
To self-consistently evolve the plasma and its electromagnetic fields, the basic electromagnetic PIC loop performs four steps repeatedly until the end of the simulation: interpolate the electromagnetic fields acting on each plasma macro-particle from the grid, advance the macro-particle position and momentum in time using the resulting Lorentz force, advance the electromagnetic fields in time by integrating Maxwell's equations and finally, project the current and charge density contribution of each macro-particle onto the grid. Derivations of these steps can be found in \cite{BirdsallLangdon2005,Arber2015,Derouillat2018}.

The described PIC loop and its extensions (e.g. ionization \cite{Nuter2011,Chen2013,Massimo2020}) can be used for a wide variety of simulations of interest for wakefield acceleration, e.g. the study of laser-driven and beam-driven injection and acceleration, beam focusing through plasma lenses \cite{Lehe2014,Marocchino2017}, the initial ionization process for the formation of HOFI channels \cite{Mewes2023,Miao2024}). In principle many experimental conditions, such as distributions for the laser fields, can be incorporated allowing for a high quantitative agreement with experimental data \cite{Ferri2016,Moulanier2023a,Moulanier2023b,Pierce2023}. 

Two-dimensional Cartesian simulations of wakefield acceleration do not provide quantitatively accurate results \cite{Davoine2008}. However, a full 3D PIC modeling of these studies can become easily unfeasible, even with the advanced parallelization techniques used to parallelize the PIC algorithm. In particular, in laser-driven acceleration the huge disparity between the largest space-time scale to simulate (the acceleration length, of the order of at least one millimetre) and the smallest space-time scale to resolve (the laser wavelength, of the order of a micron) is one of the main factors increasing the computational size of the problem.

For this reason, techniques to reduce the size of the problem based on physical considerations are often used, e.g. the use of cylindrical geometry with azimuthal mode decomposition \cite{Lifschitz2009}, the laser envelope / ponderomotive guiding center/ time-averaged ponderomotive approximation \cite{Benedetti2010,Cowan2011, Mora1997, Gordon2000, Benedetti2018,Massimo2019,Terzani2019,Massimo2019cylindrical,OsirisPGC2020,Terzani2021}, the Lorentz boosted frame technique \cite{vay2007noninvariance,Yu2016}, the quasi-static approximation \cite{Mora1997, Huang2006, Tomassini2016MatchingSF, Li2022, HiPACE2022,Wang2022QSA,AdvancedQSA2022}, and the hybrid fluid-kinetic approach to describe the plasma electrons \cite{Mora1997,Benedetti2010,Paradkar2013,Tuckmantel2014,Sosedkin2016,Massimo2016,Ferranpousa2019,Bexevanis2018,Golovanov2023}. Many of these techniques can also be coupled, e.g. using the cylindrical geometry with the quasi-static approximation and a laser envelope model \cite{Mora1997}. However, some physical phenomena cannot be accurately described with specific techniques, e.g. self-injection cannot be modeled by a purely quasi-static approximation.

The state-of-the art parallelization strategy to parallelize PIC simulations consists in decomposing the physical domain into subdomains (including the macro-particles and fields contained there), which are distributed to different computing units working in parallel. Given the non-uniformity of the macro-particle distribution in wakefield acceleration (especially in nonlinear regimes) and that the macro-particle operations represent the main computational load, advanced techniques are used to improve the performances of PIC codes, e.g. dynamic load balancing of the macro-particle load \cite{Beck2016,Miller2021}, Single Instruction Multiple Data (SIMD) vectorization of the macro-particle operations \cite{Beck2019}, macro-particle merging \cite{Vranic2015}, task parallelization \cite{Guidotti2021,Massimo2022}, multi-level mesh refining \cite{Innocenti2013,Beck2014,Fedeli2021} and different domain decomposition for communication of macro-particles and fields on the grid \cite{Derouillat2020}.

Examples of PIC codes implementing these techniques (each with a unique set of compatible features) include, but are not limited to, the following: ALaDyn \cite{ALaDyn2008}, Architect \cite{Massimo2016}, CALDER/CALDER-CIRC \cite{Lefebvre_2003,Lifschitz2009}, EPOCH \cite{Arber2015}, FBPIC \cite{Lehe2016}, Hipace++ \cite{HiPACE2022}, LCODE \cite{Sosedkin2016}, Osiris\cite{Fonseca2002}, QFluid \cite{Tomassini2016MatchingSF}, QPAD \cite{Li2022}, QuickPIC \cite{Huang2006}, Smilei \cite{Derouillat2018}, VLPL/H-VLPL \cite{Pukhov1999}, WAKE/WAKE-EP \cite{Mora1997,Paradkar2013}, Wake-T \cite{Ferranpousa2019}, WAND-PIC \cite{Wang2022QSA}, WarpX \cite{Fedeli2021}. References to studies of interest for EuPRAXIA using some of these codes can be found in the literature \cite{Nghiem_2019}. 

\subsubsection{\label{ssec:tools-mps}Multi-Physics Simulations}
Despite the methods listed above to reduce problem size and parallelize operations, there exists a variety of problems for which PIC codes are not suited due either to the physical size of the problem being considered or the timescales on which the plasma evolution must be simulated. Multi-physics simulation tools can be employed to simulate plasma evolution in the entire plasma source over nanosecond to microsecond timescales. Examples of cases here are the ionization and evolution of plasma in a capillary discharge source and the formation and evolution of an HOFI plasma channel \cite{Mewes2023}. 
In addition, for gas target development towards drivers in the $>100$W average power range, the design of the targets requires combining a multitude of physics from computational flow dynamics to heat transfer physics and laser ablation models etc.. This kind of problem would seem to be one well suited to the use of multi-physics codes based on finite-element methods. 

\section{Sustainability}


The rapid development and widespread use of conventional accelerator technology comes at a high environmental cost: accelerators are inherently power-hungry, they contribute significantly to atmospheric CO$_2$ emissions, and they rely heavily on rare earths or other materials that require extensive and aggressive mining. In light of the critical climate emergency that we are currently facing, developments in this area are thus carefully scrutinised by the public, NGOs, and governments, which are asking relevant questions on the environmental sustainability of such technology (see, e.g., Refs. \cite{Seidler2021, Owen2022, Strive2023}). 
Large-scale conventional accelerator facilities in Europe and elsewhere (e.g., CERN, DESY, and Fermilab) are starting to respond to this issue with the establishment of specific sustainability offices and the publication of self-assessment documents and roadmaps \cite{ReportDirectorFermilab2019, DESYSustainability2022, CERNEnvironmentReport2023}. It is widely expected that plasma-based accelerator systems will present several key advantages in this area with preliminary indications that they can operate at a reduced environmental cost. For example, plasma-based accelerators offer the potential to operate at much lower rates of electrical power consumption owing in part to their compactness. 
Additionally, plasma-based accelerators also predominantly rely on materials of low environmental impact (such as low-Z gases like hydrogen and helium) and are significantly less dependent on the mining of rare-earth elements. 
But, as the repetition rates for the plasma accelerator drivers are increasing, new steps will be required, such as employing closed-loop gas systems that can recycle the gas in the accelerator and increasing the efficiency of laser drivers. 

\section{\label{sec:conclusions}Conclusions}

Plasma components and systems are the key elements of future EuPRAXIA facilities, and their design, manufacture and optimization is a challenge that combines extensive knowledge of fluid dynamics, materials engineering and plasma physics. 
While different technical approaches and solutions will be required for the beam and the laser-driven cases, there is a significant amount of fundamental overlap. 
All plasma components and systems begin with the supply of a neutral material, usually but not always gas, to an interaction region. 
The properties of the plasma can be manipulated either by tailoring the neutral material or by hydrodynamic shaping of the plasma itself. This aspect of control is of paramount importance to the delivery of high-quality electron beams for applications. 
Additionally, several common challenges exist for all plasma components and systems listed here. These include high average power operation for extended periods, integration of plasma components and systems into the vacuum and beamline infrastructures, material robustness and longevity under extreme conditions, plasma component metrology and plasma component simulations.
In addressing these challenges, further standardization of plasma components and systems as well as the simulation tools and diagnostics used to understand them, will be of benefit.

In many cases, the requirements on plasma components and systems envisaged for EuPRAXIA go beyond the state of the art available today. However, it is also clear that there has been consistent and rapid progress in all areas of interest and that the technological advancements required should be achievable on a timescale commensurate with the development of the EuPRAXIA facilities. As such, a roadmap for plasma source development would be beneficial, to help identify key challenges and a wholistic approach to tackling them. 









\providecommand{\noopsort}[1]{}\providecommand{\singleletter}[1]{#1}%

\end{document}